# Commensurate 4a$_0$-period Charge Density Modulations throughout the Bi$_2$Sr$_2$CaCu$_2$O$_{8+x}$ Pseudogap Regime

A. Mesaros[a], K. Fujita[b], S.D. Edkins[a,c], M.H. Hamidian[d,e], H. Eisaki[f], S. Uchida[f,g], J.C. Séamus Davis[a,b,c,h], M. J. Lawler[a,i] and Eun-Ah Kim[a]

[a]  LASSP, Department of Physics, Cornell University, Ithaca, NY 14853, USA.
[b]  CMPMS Department, Brookhaven National Laboratory, Upton, NY 11973, USA.
[c]  School of Physics and Astronomy, University of St. Andrews, St. Andrews, Fife KY16 9SS, Scotland
[d]  Department of Physics, Harvard University, Cambridge, MA 02138, USA
[e]  Department of Physics, University of California, Davis, Davis, CA 95616, USA.
[f]  Institute of Advanced Industrial Science and Technology, Tsukuba, Ibaraki 305-8568, Japan.
[g]  Department of Physics, University of Tokyo, Bunkyo-ku, Tokyo 113-0033, Japan.
[h]  Tyndall National Institute, University College Cork, Cork T12R5C, Ireland.
[i]  Department of Physics, Binghamton University, Binghamton, NY 13902-6000, USA.

**ABSTRACT  Theories based upon strong real space ($r$-space) electron-electron interactions have long predicted that unidirectional charge density modulations (CDM) with four-unit-cell ($4a_0$) periodicity should occur in the hole-doped cuprate Mott insulator (MI). Experimentally, however, increasing the hole density $p$ is reported to cause the conventionally-defined wavevector $Q_A$ of the CDM to evolve continuously as if driven primarily by momentum-space ($k$-space) effects. Here we introduce phase-resolved electronic structure visualization for determination of the cuprate CDM wavevector. Remarkably, this new technique reveals a virtually doping independent locking of the local CDM wavevector at $|Q_0| = 2\pi/4a_0$ throughout the underdoped phase diagram of the canonical cuprate Bi$_2$Sr$_2$CaCu$_2$O$_8$. These observations have significant fundamental consequences because they are orthogonal to a $k$-space (Fermi-surface) based picture of the cuprate CDM but are consistent with strong-coupling $r$-space based theories. Our findings imply that it is the latter that provide the intrinsic organizational principle for the cuprate CDM state.**

*CuO$_2$ pseudogap, Commensurate Charge Density Modulation,  Phase Discommensuration*

*Introduction*

Strong Coulomb interactions between electrons on adjacent Cu sites results in complete charge localization in the cuprate Mott insulator (MI) state[1]. When holes are introduced, theories based upon the same strong *r*-space interactions have long predicted a state of unidirectional modulation of spin and charge[2-11], with lattice-commensurate periodicity for the charge component. Experimentally, it is known that even the lightest hole doping of the MI state produces nanoscale clusters of CDM[12,13] which implies immediately that *r*-space interactions predominate. However, with increasing hole density $p$, the conventionally-defined wavevector $\boldsymbol{Q_A}$ of the CDM is reported to increase[14] or diminish[15] continuously as if driven primarily by *k*-space (Fermi surface) effects. Distinguishing between the *r*-space and *k*-space theoretical perspectives is critical to identifying the correct fundamental theories for the phase diagram and Cooper pairing mechanism in underdoped cuprates. Here we introduce a new approach to this challenge by applying phase-resolved electronic structure visualization[16,17,18] in combination with the new technique of phase demodulation-residue minimization, to explore the CDM wavevector. Using these methods, we visualize the phase discommensurations[19] and their influence on the doping dependence of both the conventionally-defined $\boldsymbol{Q_A}$ and the fundamental local wavevector $\boldsymbol{Q_0}$ of the underdoped cuprate CDM state.

*Charge Density Modulations in the Pseudogap Phase*

**1** As holes are introduced into the CuO$_2$ plane of the MI, the first non-magnetic state to appear is the pseudogap (PG) phase (Fig. 1*A*). It contains nanoscale CDM clusters[12,13] even at lowest hole-density $p$, near $p$=0.06, these CDM clusters percolate and the superconducting state appears[12]. X-ray scattering experiments[15] now report a robust CDM state throughout the range 0.07<$p$<0.17 spanning the pseudogap regime. Both the PG and CDM states terminate somewhere near $p$~0.19 and give way to a simple d-wave superconductor. A fundamental reason for the great difficulty in understanding this complex phase diagram has been inability to discern the correct theoretical starting point. Should one focus on the intense *r*-space electron-electron interactions that form the basis for the parent MI state? Or, should one focus upon a Fermi surface of momentum space eigenstates representing delocalized electrons?

**2** A new opportunity to address these questions has emerged recently, through studies of the CDM phenomena now widely observed in underdoped cuprates[15,20,21]. Pioneering studies of CDM in $La_2BaCuO_4$ and $La_2NdSrCuO_4$ near $p=0.125$ discovered charge modulations of period $4a_0$ or $\boldsymbol{Q}=((1/4,0);(0,1/4))2\pi/a_0$ (Refs. 14, 21). The initial intuitive explanation for a periodicity that was half that predicted from a Hartree-Fock momentum space treatment, invoked an *r*-space model involving local magnetic moments whose antiferromagnetic order becomes frustrated upon hole-doping. A variety of powerful theoretical techniques[2-11] support this strongly interacting *r*-space viewpoint. In the interim, however, CDM phenomena have been discovered within the pseudogap regime of many other underdoped cuprates[15,20,21]. In these studies, the magnitude of the conventionally-determined CDM wavevector $\boldsymbol{Q}$ is reported to increase/diminish with increasing $p$, as if evolution of momentum space electronic structure with carrier density is the cause. Thus, distinguishing between an *r*-space or Fermi-surface-based theoretical approach to the cuprate CDM remains an outstanding and fundamental challenge, and one that is key to the larger issue of controlling the balance between different electronic phases. The reason is that, in the *k*-space context[22-25], competition for spectral weight at the Fermi surface between different electronic states including the superconductivity is a zero sum game: suppressing one state amplifies another and vice versa. By contrast, in the strong interaction *r*-space context[2-11], the physics of holes doped into an antiferromagnetic MI yields "intertwined" states[4,8-11,26,27] including superconductivity, where closely related ordered-states are generated simultaneously by the same microscopic interactions.

### *Charge Density Modulations and Phase Discommensurations*

**3**     Understanding the cuprate CDM phenomenology has proven challenging[28-30] because its ***q***-space peaks are typically broad with spectral weight distributed over many wavevectors[15-21], and also because of the novel form factor symmetry[17,32,33]. For example, Figs 1B,C shows a typical image of the electronic structure of underdoped $Bi_2Sr_2CaCu_2O_8$. $\psi_R(r)$; here $\psi_R(r, 150\ meV) \equiv I(r, 150\ meV)/I(r, -150\ meV)$ and $I(r, V)$ is the measured tunnel current at position ***r*** for bias voltage V. The CDM exhibits a *d*-symmetry form factor meaning that modulations at the *x*-axis oriented planar oxygen sites $O_x(r)$ are π out of phase with those at *y*-axis oriented oxygen sites $O_y(r)$, as shown schematically by the overlay to Fig. 1*B*. Thus, the complex Fourier transforms $\tilde{O}_x(q); \tilde{O}_y(q)$ of sublattice-resolved images $O_x(r); O_y(r)$ that are derived[17,33] from $\psi_R(r)$, yield the d-symmetry form factor Fourier amplitudes $|\tilde{\psi}_R(q)| = |\tilde{O}_x(q) - \tilde{O}_y(q)|$ shown in Fig. 1D (see *SI Text* section 1). One sees directly the wide range of ***q*** values that exist under each CDM peak in $\tilde{\psi}_R(q)$ (dashed boxes Fig. 1*D*). Such broad peaks indicate quenched disorder of the CDM but with two quite distinct possibilities for the identity of the fundamental ordered state: (i) an incommensurate CDM state whose wavevector ***Q*** evolves continuously along with the Fermi wavevector ***k**$_F$* (e.g. Fig. 1*E*) but is perturbed by disorder or, (ii) a commensurate CDM with constant fundamental wavevector ***Q**$_0$* driven by strong-coupling ***r***-space effects (e.g. Fig. 1*F*), but whose wavevector defined at the maximum of the fitted Fourier amplitudes, ***Q**$_A$*, evolves due to changing arrangements of discommensurations (DC).

**4**     McMillan defined[19] a 'discommensuration' as a defect in a commensurate CDM state where the phase of the CDM jumps between discrete lattice-locked values. For example, consider a sinusoidal modulation in one spatial dimension with a commensurate period $4a_0$

$$\psi(x) \equiv A \exp[i\Phi(x)] = A \exp[i(Q_0 x + \varphi)] \tag{1}$$

Here $Q_0 = \frac{2\pi}{4a_0}$ is the commensurate wavevector, $A$ is the amplitude, $\Phi(x) = Q_0 x + \varphi$ is the position-dependent phase argument of the function $\psi(x)$. To form phase-locked regions, the phase off-set can take one of four discrete values $\frac{\varphi}{2\pi} = 0, \frac{1}{4}, \frac{2}{4}, \frac{3}{4}$. The DCs then form boundaries between these regions as indicated by different colors in Figs 2*Ai*, 2*Aii*. When a commensurate CDM (Eqn. 1) is frustrated by Fermi-surface-based tendencies, a regular DC array allows more/fewer modulations to be accommodated through successive jumps in phase (Fig. 2*Aii* and Ref. 19). The result is a new phase-averaged wavevector $\bar{Q}$ that depends on the profile of the DC array $\varphi(x)$ through $Q_0 x + \varphi(x) \equiv \bar{Q}x + \tilde{\varphi}(x)$, where the slope $\bar{Q}$ is chosen so that the residual phase fluctuations $\tilde{\varphi}(x)$ average to zero ($\overline{\tilde{\varphi}(x)} = 0$). Graphically, finding the $\bar{Q}$ is then equivalent to finding the best linear function for $\Phi(x)$ as shown in Fig. 2*Aii*. In this case, the difference in slope between the commensurate and phase-averaged wavevectors, $Q_0$ and $\bar{Q}$, is called the incommensurability $\delta \equiv \bar{Q} - Q_0$ (Fig. 2*Aiii*) of such nominally incommensurate phase. Note that such a DC array does not affect the correlation length of the CDM even though it does shift the Fourier amplitude defined wavevector $Q_A$ to value $\bar{Q}$ from the fundamental commensurate wavevector $Q_0$ (Fig. 2*Aiii*). In contrast, when the combined phase jumps of all the DCs average to zero (Figs 2*Bi*, 2*Bii*) the phase-averaged wave vector $\bar{Q}$ equals the local commensurate wave vector, i.e. $\bar{Q} = Q_0$. Here, in the absence of additional amplitude disorder, the $Q_A$ should also peak at $\bar{Q} = Q_0$ (Fig. 2*Biii*). However, in the most realistic case where disorder in the CDM amplitude and the random spatial arrangement of DCs coexist, $Q_A$ is demonstrably a poor measure of the fundamental commensurate wavevector $\bar{Q} = Q_0$ (Fig. 2*Ci*, *Cii* and SI Text sections 2 and 3).

**5** How then can one correctly determine the fundamental $Q_0$ of the CDM in underdoped cuprates? The spatial arrangement of DCs are inaccessible to diffraction probes designed to yield the Fourier amplitude of the CDM, although the situation in Fig. 2Aiii may be detectable through the observation of satellite peaks at $\bar{Q} \pm \delta$ (SI Text sections 2 and 5). Phase sensitive transmission electron microscopy can achieve DC visualization[34] but has never been used on cuprates. Instead, we consider CDM visualization using spectroscopic imaging scanning tunneling microscopy[20] because it offers full access to *both* the amplitude and phase of $\tilde{\psi}(q)$, with the definition of phase referenced to the underlying atomic lattice (Ref. 33). Then, based on such phase visualization capabilities, we introduce a new approach for identifying the fundamental wavevector of the cuprate CDM state. To achieve what is graphically represented in Fig. 2 as the dashed linear-fit to a measured phase profile $\Phi(x)$, we use an algorithmic procedure that evaluates the demodulation of measured CDM image $\psi(r)$ at each possible wavevector $q$ using the *demodulation residue*

$$R_q[\psi] \equiv \int \frac{dx}{L} Re[\psi_q^*(x)(-i\partial_x)\psi_q(x)]. \qquad (2)$$

Identifying the value of wavevector $q$ for which $|R_q[\psi]|$ is a minimum is the 2-dimensional equivalent to finding the best-fit slope to $\Phi(x)$ in Fig. 2*A*, *B*. The resulting wavevector $q = \bar{Q}$ with high accuracy (see *SI Text* section 2), and determination of this $\bar{Q}$ is the general objective and utility of this new technique.

### *Phase-resolved Imaging and Phase Demodulation Residue Analysis*

*6* We apply demodulation residue analysis to study two-dimensional short-range ordered CDM images typical of underdoped $Bi_2Sr_2CaCu_2O_8$, e.g. Fig. 3*A* at $p$=0.06. Figure 3*B* shows the d-symmetry form factor Fourier amplitude $|\tilde{\psi}(q)|$ derived from $\psi(r)$ in Fig. 3*A*. The amplitude and phase of modulations in $\psi(r)$ can be decomposed into two unidirectional components along *x*, *y*. We define the demodulation residue $R_q$ for each trial $q$ (see *SI Text* section 4) over a wide range in the Fourier space inside the dashed box in Fig. 3B, and use it as phase-sensitive metric for deciding how close each $q$ is to the phase-averaged wavevector $\overline{Q}$. In the ideal limit, a long-range ordered CDM with wave vector $\overline{Q}$ will have zero demodulation residue, i.e., $|R_{\overline{Q}}| = 0$ and the Fourier amplitude will vanish for $q \neq \overline{Q}$. However, the measured Fourier amplitude distribution is typically broad and asymmetric (Fig. 3B and see Fig.3.7b of Ref.20) and not well-fitted by a smooth peak (Fig. 3C and *SI Text* section 3). Hence we will seek the minimum of $|R_q|$ for which $\tilde{\psi}(q)$ retains an appreciable amplitude. For the data shown in Fig. 3A, B we calculate the $|R_q[\psi(q)]|$ for every pixel identified by a colored symbol in Fig. 3E (see *SI Text* section 4). In Fig. 3E we plot the value of both $|R_q[\psi(q)]|$ and the amplitude in the dFF Fourier transform $\tilde{\psi}(q)$ for each of these pixels in Fig 3D. This shows that the procedure singles out one wavevector for the CDM in the y-direction with a nearly vanishing demodulation residue, a gap between this $|R_q|$-minimizing $q$ (which we denote by $\overline{Q}_Y$) and the rest of the wavevectors, and that this occurs for wavevector within the Fourier intensity peak. The $\overline{Q}_Y$ is identified as the green pentagon within black square box in Fig 3D. Instructively this demodulation residue-minimizing $q = \overline{Q}_Y$ does not equal the wavevector at which the Fourier amplitude $|\tilde{\psi}(q)|$ is the largest. Indeed the power of the $|R_q|$-minimization approach is that it singles out the phase-averaged $\overline{Q}_Y$ for this CDM, from a broad and asymmetric Fourier amplitude peak where $Q_A$ is unreliable (see *SI Text* section 2 and 3). Most remarkably, we find that $|\overline{Q}_Y|$ has commensurate value $\frac{2\pi}{4a_0}$ within the error. Moreover, imaging the phase of CDM at the commensurate $\overline{Q}_Y$ reveals in $r$-space where CDM phase is locked to the four expected discrete values ($n\frac{2\pi}{4}; n = 0,1,2,3$, see Fig. 3F), forming locally commensurate $Q_0 = \frac{2\pi}{4a_0}$ regions of the fundamental CDM (see *SI Text* section 5). In this highly typical case of a $Bi_2Sr_2CaCu_2O_8$ $\psi(r)$, the phase slips of the discommensurations average to zero as in Fig. 2B, confirming the fundamental $Q_0 = \overline{Q}_Y$.

**7**     Given this demonstrated capability of $|R_q[\psi(q)]|$ minimization to extract the defining $Q$ from short-range CDM data, we next turn our attention to the doping dependence of fundamental $Q_X$; $Q_Y$ throughout the pseudogap regime of underdoped Bi$_2$Sr$_2$CaCu$_2$O$_8$ (*SI Text* section 6). Figure 4A contains two side by side panels; the left shows measured $|\tilde{\psi}(q)|$ while the right is the measured $|R_q[\psi(q)]|$ analysis for its y-axis modulations. Figure *4A-E* then show a series of such pairs of measured $|\tilde{\psi}(q)|$) and $|R_q[\psi(q)]|$ for five different Bi$_2$Sr$_2$CaCu$_2$O$_8$ hole-densities *p*=0.06, 0.08, 0.10, 0.14, 0.17. In all cases, the demodulation residue-minimizing process clearly singles out the minimized values in $|R_q[\psi(q)]|$ for the phase-averaged CDM wavevectors. This is evident in the sharp minimum that is observed near the (0,0.25)2π/a$_0$ point (marked by a cross in each right-hand $|R_q[\psi(q)]|$ panel). Therefore, the most striking result as summarized in Fig. *4F* is that the measured magnitudes of the average wavevectors $\overline{Q}_X$; $\overline{Q}_Y$ of the Bi$_2$Sr$_2$CaCu$_2$O$_8$ CDM are all indistinguishable from the lattice commensurate values $Q_0 = (0,1/4)2\pi/a_0; (1/4,0)2\pi/a_0$ making the fundamental wavevectors $Q_X$; $Q_Y$ equal to $Q_0$ and virtually doping independent (*SI Text* section 6). Moreover, the largest deviation of the conventional amplitude-derived $Q_A$ from the phase-optimized value $\overline{Q} = Q_0$ is at lowest doping, which can be associated with the observed higher density of discommensurations at the same doping (SI Text section 2).

## *Ubiquity of Lattice-Commensurate Charge Density Modulations*

**8**     Comparison of this result with reports of a preference for a CDM periodicity of *4a$_0$* in YBa$_2$Cu$_3$O$_{7-x}$-based heterostructures[35], in the NMR-derived view of the lattice-commensurate CDM in YBa$_2$Cu$_3$O$_{7-x}$,[36] and in the pair density wave state of Bi$_2$Sr$_2$CaCu$_2$O$_8$[37], point to growing evidence for a unified phenomenology of lattice-commensurate CDM across disparate cuprate families. Of course, the wavevectors $Q_A$ of maximum intensity in X-ray diffraction patterns for YBa$_2$Cu$_3$O$_{7-x}$ and La$_2$Sr(Ba)CuO$_4$ families evolve continuously with doping and appear generally incommensurate[14,15].

However, discommensuration configurations of the type in Fig. 2*A* will result in such an incommensurate average wavevector $\overline{Q} = Q_A$ even though the fundamental wavevector $Q_0$ of the CDM is commensurate, so that evolution of cuprate DC arrays (e.g. Fig. 2) can yield the incommensurate wavevector evolution detected by X-ray scattering (*SI Text* section 7); a related hypothesis has long been considered[38]. Our application of the classic theory of CDM discommensuration disorder[19] (Fig. 2) with novel CDM phase-resolved imaging and analysis, reveals this as the correct picture for $Bi_2Sr_2CaCu_2O_8$. This motivates the hypothesis that doping dependence of $Q_A$ across all cuprate families is caused by a competition between incommensurate modulations promoted by evolving Fermi arcs and a lattice-commensurate CDM state, with this competition being resolved through DC configurations.

**9**   The $|R_q|$-minimization technique introduced here is designed to utilize the additional CDM-information available only from phase-resolved imaging[16,17,33]. Remarkably, it reveals a doping independent locking of the phase-averaged CDM wavevector to a lattice commensurate wavevector $|Q_0| = 2\pi/4a_0$ oriented with the Cu-O-Cu bonds in $Bi_2Sr_2CaCu_2O_8$ (Fig. 4). Moreover, we directly detect the CDM discommensurations between phase locked commensurate regions which generate this situation (Fig. 3*F*). These observations have significant fundamental consequences for understanding the mechanism of the cuprate CDM state. They are orthogonal to a weak-coupling *k*-space based picture for CDM phenomena, in which the fundamental wavevector should increase or decrease monotonically with doping, or should evolve in discrete jumps even with 'lattice locking'[39]. By contrast, a lattice-commensurate CDM state has been obtained comprehensively in different types of strong-coupling *r*-space based theories[2-11]. For underdoped $Bi_2Sr_2CaCu_2O_8$ at least, our data are far more consistent with such lattice-commensurate strong-coupling *r*-space theories being the intrinsic organizational principle for the cuprate CDM phenomena. Furthermore, nanoscale clusters of lattice-commensurate CDM are the first broken-symmetry state to emerge at lightest hole-doping[12,13], multiple transport and spectroscopic measurements of cuprate quasiparticles have recently been demonstrated quite consistent with lattice-commensurate $r$-space theories,[40] and $YBa_2Cu_3O_{7-x}$ NMR studies[36,41] are also most consistent with them. Explorations of universality of lattice-commensurability of CDMs in other cuprate compounds can now be pursued using these phase-resolved imaging and $|\boldsymbol{R_q}|$-minimization techniques.

**Figure Captions**

**Fig. 1. CHARGE DENSITY MODULATIONS IN CUPRATES**

(A) Schematic phase diagram of hole-doped cuprates. The high-temperature superconductivity coexists with d-symmetry form factor charge modulations (compare blue and pink "domes", respectively) through most of the underdoped regime.

(B) Typical measured $\psi_R(r, 150\text{ meV})$ image $Bi_2Sr_2CaCu_2O_8$ in the charge modulation phase. The subatomic resolution image shows a typical charge modulation pattern of d-symmetry form factor, spanning 8 lattice constants horizontally (compare to *E*, *F*), and with overlay showing how the d-symmetry form factor affects each oxygen site. The green crosses mark positions of Cu atoms in the underlying $CuO_2$ plane.

(C) Typical $\psi_R(r, 150\text{ meV})$ of underdoped $Bi_2Sr_2CaCu_2O_8$; the short range nature of the charge modulations is clear.

(D) The d-symmetry form factor Fourier amplitudes $|\tilde{\psi}_R(q, 150\text{meV})|$ calculated using complex Fourier transforms of sublattice-resolved images $O_x(r); O_y(r)$ derived from *C*.

(E) Modeled d-form factor charge density wave which represents an incommensurate modulation by having horizontal wavevector of length $Q = 0.311 \times \frac{2\pi}{a}$, where $a$ is lattice unit. The density values are sampled and color-coded at each Cu site (green crosses), O site, and center of $CuO_2$ plaquette, to emphasize the modulation pattern and relation to underlying lattice. The initial phase is chosen so that the modulation on leftmost line of Cu sites matches in value the commensurate modulation in

(F) Incommensurate modulations such as the one shown naturally arise from Fermi surface instabilities, and therefore have the Fermi surface nesting wavevector $Q = 2k_F$ and period $2\pi/Q = \pi/k_F$. The dashed line is the profile of density wave along horizontal direction, without imposing a d-form factor intra-unit-cell structure, and period marked by length of double arrow.

(G) Modeled d-form factor charge density wave which is commensurate, having wavevector $Q = \frac{1}{4} \times \frac{2\pi}{a}$ and period $2\pi/Q = 4a$ with $a$ the lattice unit. The density values are sampled and color-coded at each Cu site (green crosses), O site, and center of $CuO_2$ plaquette, to emphasize the modulation pattern and relation to underlying lattice. The initial wave phase is chosen so that the modulation maximum occurs on a horizontal O site. The dashed line is the profile of density wave along horizontal direction and period marked by length of double arrow.

**Fig. 2. DISCOMMENSURATIONS, FUNDAMENTAL AND AVERAGE WAVEVECTORS**

Panels Ai-Aiii show a discommensuration model in a situation which may apply to YBCO. Model in panels B, C corresponds to our findings in BSCCO.

(A*i*) Modulation (blue, thick) is the real part of complex wave $\psi(x) = e^{i(Q_0 x + \varphi(x))}$ having commensurate domains with local wave vector $Q_0 = \frac{1}{4} \times \frac{2\pi}{a}$ (period $4a_0$). Colors (see legend) label the modulation phase within the domains, determining the alignment of modulation maxima (labeled 1…10) and underlying lattice. Phase slips, occurring at discommensurations between domains, each of size $\pi$, add up to give an average $\bar{Q} = 0.3 \times \frac{2\pi}{a}$, so that 10 modulation maxima are squeezed into $31a$.

(A*ii*) The phase argument $\Phi(x) = Q_0 x + \varphi(x)$ of $\psi(x)$ in A*i*. Commensurate domains occur in regions (colored) where $\Phi(x)$ is parallel to line $Q_0 x$ (red dashed line). The average $\bar{Q} = 0.3 \times \frac{2\pi}{a}$ is seen as slope of best-fit line to $\Phi(x)$ (blue dashed line). The difference in slope gives the incommensurability $\delta = \bar{Q} - Q_0$.

(A*iii*) Fourier amplitudes $|\tilde{\psi}(q)|$ of the modulation $\psi(x)$ in A*i* (blue line) show singular peaks starting at $\bar{Q} = 0.3 \times \frac{2\pi}{a} \neq Q_0$ with satellites separated by $2\delta$, since discommensurations of size $\pi$ form a periodic array. The satellites depend on DC profile and are sensitive to disorder (see *SI Text* section 2). The phase-sensitive figure of merit, demodulation residue $|R_q|$ (red dashed line), as a function of $q$ has the minimum exactly at the average $\bar{Q}$. By definition its minimum corresponds to slope of best fit line through $\Phi(x)$ (see *Aii*).

(B*i*) Modulation (blue, thick) is the real part of complex wave $\psi(x) = e^{i(Q_0 x + \varphi(x))}$ having commensurate domains with fundamental local wave vector $Q_0 = \frac{1}{4} \times \frac{2\pi}{a}$ (period $4a_0$). Colors label the modulation phase within the domains, determining the alignment of modulation maxima (labeled 1...8) and underlying lattice. All the phase slips, of sizes $+\pi, -\frac{3\pi}{2}, +\frac{\pi}{2}$, occurring at discommensurations between domains, cancel to give an average $\bar{Q} = Q_0$, seen in preserving the 8 modulation maxima within $31a$.

(B*ii*) The phase argument $\Phi(x) = Q_0 x + \varphi(x)$ of $\psi(x)$ in *Bi*. Commensurate domains occur in regions (colored) where $\Phi(x)$ is parallel to line $Q_0 x$ (blue dashed line). The average $\bar{Q} = Q_0$ is seen as slope of best-fit line to $\Phi(x)$ which coincides with dashed line.

(B*iii*) Fourier amplitudes $|\tilde{\psi}(q)|$ of the modulation $\psi(x)$ in *Bi* (blue line) have a sharp peak at $\bar{Q} = Q_0$ and additional irregularly distributed weight due to disorder in DC's position (see Methods Section II). The calculated phase-sensitive figure of merit, demodulation residue $|R_q|$ (red dashed line), as a function of $q$ has the minimum exactly at the average $\bar{Q}$. By definition its minimum corresponds to slope of best fit line through $\Phi(x)$ (see *Bii*).

(C*i*) Modulation (blue, thick) is the real part of complex wave $\psi(x) = A(x)e^{i(Q_0 x + \varphi(x))}$ having commensurate domains with local wave vector $Q_0 = \frac{1}{4} \times \frac{2\pi}{a}$, additional smooth disorder in phase of up to $\pi/10$, and smooth disorder in amplitude $A(x)$ (details in *SI Text* section 2). All random phase slips cancel to give a $\bar{Q} = Q_0$, akin to case in panels *B*.

(C*ii*) Fourier amplitudes $|\tilde{\psi}(q)|$ of modulation $\psi(x)$ over $125a_0$ range (exemplified in *Ci*). Broad asymmetric amplitude peak makes the $Q_A$ (orange vertical line), at maximum of Lorentzian fit to amplitude (orange dashed line, multiplied by 1.5 for visibility), different than phase-averaged $\bar{Q}$ (red vertical line) at minimum of demodulation residue $|R_q|$ (red dashed line), see Methods Section II.

Fig. 3.  PHASE RESIDUE MINIMIZATION ANALYSIS OF SISTM DATA

(A) Typical measured $\psi(r)$ of Bi$_2$Sr$_2$CaCu$_2$O$_8$ in the charge modulation phase at hole doping level $p$=0.06. The subatomic resolution image shows charge modulations at pseudogap energy. Coordinate axes $x, y$ correspond to copper-oxide lattice principal axes.

(B) The Fourier transform amplitudes of d-symmetry form factor, $|\tilde{\psi}(q)|$, extracted from the measurement in *A*. Four broad intensity distributions appear due to CM, one of them (along $x$-axis) is marked by the dashed square. The unit-cell constant, $a$, is determined by Bragg peaks (red crosses).

(C) Measured d-form factor Fourier amplitudes $|\tilde{\psi}(q)|$ (dots) along $q_x$-axis in dashed square of *B*, i.e., surrounding the $Q_X$ CM peak, from origin towards Bragg peak, showing d-form factor Fourier amplitudes (dots) and the cut through best fitting smooth two-dimensional peak function (see Methods Sections I, III). The fit residual at each wavevector (vertical drop from dot to fit function) is color-coded. The integers on horizontal axis count the pixels in the Fourier transform image, i.e., wavevector length on horizontal axis is measured in units $2\pi/L$, with *L* the field-of-view size in lattice units.

(D) Close-up of the Fourier amplitudes of $|\tilde{\psi}(q)|$ from *B* marked by the dashed square. The discrete set of wavevectors in $q$-space area of Fig. 3b, i.e., surrounding the $Q_Y$ CM peak, and their d-form factor Fourier amplitudes are shown. Each pixel at which $|R_q|$ is calculated (see *E*) is color-symbol coded. Commensurate value $q = (0,1/4)$ is marked by cross.

(E) Demodulation residue $|R_q|$ versus Fourier amplitude used as two figures of merit in CM period analysis, for d-symmetry form factor component measured in $Bi_2Sr_2CaCu_2O_8$ at doping *p=0.06* (data presented in Fig.3*A*, *B*). A set of wavevectors (the color-symbol coded pixels in *E*) is used in this plot showing value of demodulation residue $|R_q| = \sqrt{(R_q^x)^2 + (R_q^y)^2}$, calculated using cutoff $\Lambda = 0.08 \times \frac{2\pi}{a}$ (see Methods Section IV), versus the Fourier amplitude $|\tilde{\psi}(q)|$, for each wavevector in the chosen set. The boxed data point is for the discrete wavevector value that is identified as minimizing the demodulation residue and therefore we define it as phase-optimized wavevector. Its length is $(0.24 \pm 0.03) \times \frac{2\pi}{a}$, where error value follows from spatial variation of residue (see *SI Text* section 4). The residue is significantly closer to zero at the phase-optimized wavevector than for others, even though there are many wavevectors having higher Fourier amplitudes. [Note that conventional analysis by fitting a broad-peak to Fourier amplitudes of this data would identify wavevector length $Q = 0.29 \times \frac{2\pi}{a}$].

(F) Measured $\psi(r)$ of $Bi_2Sr_2CaCu_2O_8$ in the charge modulation phase at hole doping level *p=0.08*. Overlay is the *x*-axis CDM demodulated by phase-optimized commensurate wavevector $|Q_X| = \frac{1}{4} \times \frac{2\pi}{a_0}$, with high amplitude having high color saturation. Phase locked domains with phase an integer multiple of $\frac{2\pi}{4}$ (color legend) are visible. Across the field of view, e.g., following the white dashed line, the phase slips between domains average to zero, evoking Fig. 2*B*.

**Fig. 4. Measured Wavevector of Charge Density Modulations throughout the Bi2212 Pseudogap Regime**

(A) Left panel shows measured $|\tilde{\psi}(q)|$ for $p=0.06$, within the square region of $q$-space (the dashed square in Fig.3B) bounded by $(0,0)$, $(\pi/2a,\pi/2a)$, $(0,\pi/a)$, $(-\pi/2a,\pi/2a)$. Right panel shows the value of demodulation residue $|R_q| = \sqrt{(R_q^x)^2 + (R_q^y)^2}$ calculated from data using cutoff $\Lambda = 0.08 \times \frac{2\pi}{a}$. The measured $|R_q|$ varies smoothly and drops quickly towards zero at a single discrete wavevector near center of image, similarly to behavior in one-dimension (see Fig. 2). The identified discrete wavevector has length $|Q| = 0.245 \times \frac{2\pi}{a}$. The position of $(0,\frac{1}{4}) \times \frac{2\pi}{a}$ is marked by a small cross. The phase-optimized wavevector $Q$ in continuous $q$-space can be at most localized within the whitest pixel area, leading to a minimal error of $0.02 \times \frac{2\pi}{a}$. The error value $0.03 \times \frac{2\pi}{a}$ is obtained from spatial variation of residue function (see *SI Text* section 4).

(B) – (E) Same analysis as shown in A, but for a series of samples with estimated hole densities $p=0.08$, $0.10$, $0.14$, $0.17$. In each case the left image shows the measured $|\tilde{\psi}(q)|$ within the square region of $q$-space which defines the considered CM orientation, e.g., for $Q_Y$ bounded by $(0,0)$, $(\pi/2a,\pi/2a)$, $(0,\pi/a)$, $(-\pi/2a,\pi/2a)$; each right image shows the measured $|R_q|$ within the square region of $q$-space marked by orange thin square on the image to its left. The corresponding position of $(0,\frac{1}{4}) \times \frac{2\pi}{a}$ is marked by a small cross. The pixel at which the $|R_q|$ is found to be a minimum is identified as the phase-optimized CM wavevector for that carrier density.

(F) The lengths of wavevectors $Q_X, Q_Y$ extracted from measured $Bi_2Sr_2CaCu_2O_8$ underdoped samples at different dopings $p$ using $|R_q|$ minimization (see *A* to *E*). The error in value of the phase-optimized wavevector is obtained from spatial variation of residue (see *SI Text* section 4), and is comparable to the error caused by discreteness of choices for $Q$. For doping $p > 0.14$ the d-form factor CM is less pronounced and there is a larger error in value of phase-optimized $Q$. Note the doping-independent trend and values consistent with commensurate value 1/4. Reported doping-dependent values of CDM wavevector length $|Q_A|$ in: BSCO (disks, colored light to dark in order for References 42, 43, 33), YBCO (squares, light to dark for: source 132 from Ref.15, Ref. 44, and Ref. 36), and LBCO (diamonds, reported in Ref. 44).

**Acknowledgements**

A.M. acknowledges support by the U.S. Department of Energy, Office of Basic Energy Sciences, Division of Materials Science and Engineering under Award DE-SC0010313; E.-A.K. acknowledges Simons Fellow in Theoretical Physics Award #392182; SDE acknowledges studentship funding from the EPSRC under Grant EP/G03673X/1; M.H.H. acknowledges support from the Moore Foundation's EPiQS Initiative through grant number GBMF4544; S.U. and H.E. acknowledge support from a Grant-in-Aid for Scientific Research from the Ministry of Science and Education (Japan) and the Global Centers of Excellence Program for Japan Society for the Promotion of Science; J.C.S.D. acknowledges gratefully the hospitality and support of the Tyndall National Institute, University College Cork, Cork, Ireland. Experimental studies were supported by the Center for Emergent Superconductivity, an Energy Frontier Research Center, headquartered at Brookhaven National Laboratory and funded by the U.S. Department of Energy under DE-2009-BNL-PM015.

**Author Contributions:** A.M., M.J.L. and E.-A. K developed the phase-error minimization techniques and protocols; M.H.H., S.D.E., and K.F. carried out the experiments; J.C.D. and E.-A.K. supervised the project and wrote the paper with key contributions from A.M., M.J.L and K.F. The manuscript reflects the contributions and ideas of all authors.

**Author Information** Reprints and permissions information is available at www.nature.com/reprints. The authors declare no competing financial interests. Readers are welcome to comment on the online version of the paper. Correspondence and requests for materials should be addressed to J.C.S.D jcseamusdavis@gmail.com and E.-A. K.; eun-ah.kim@cornell.edu

Figure1

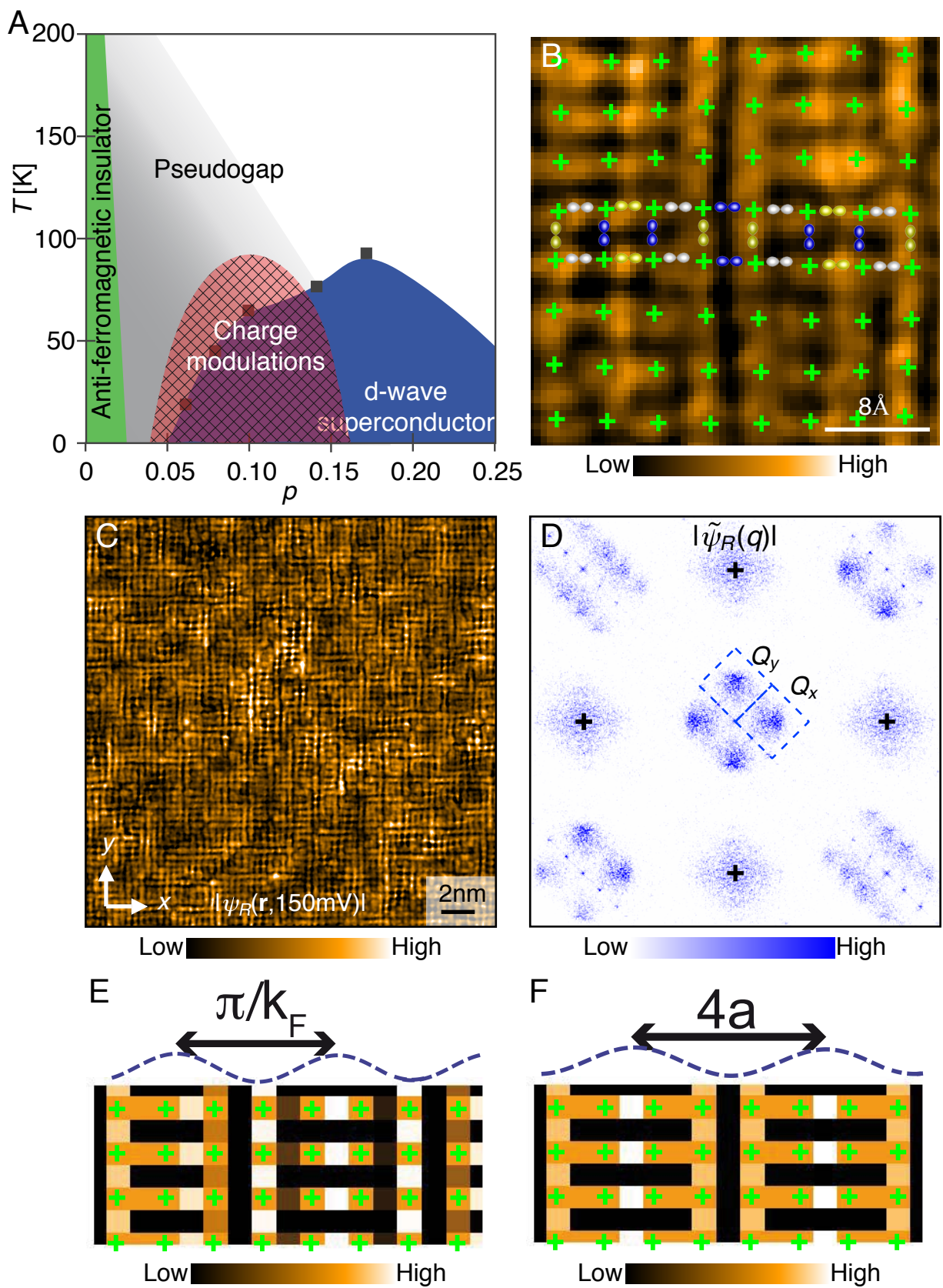

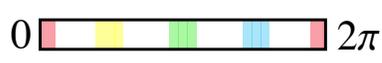

Figure 2

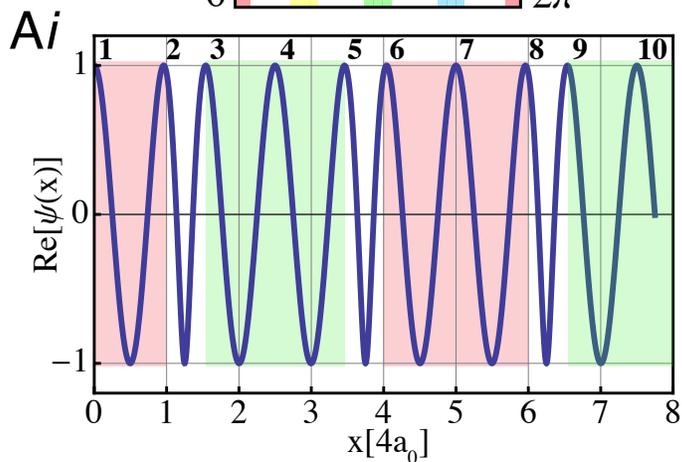
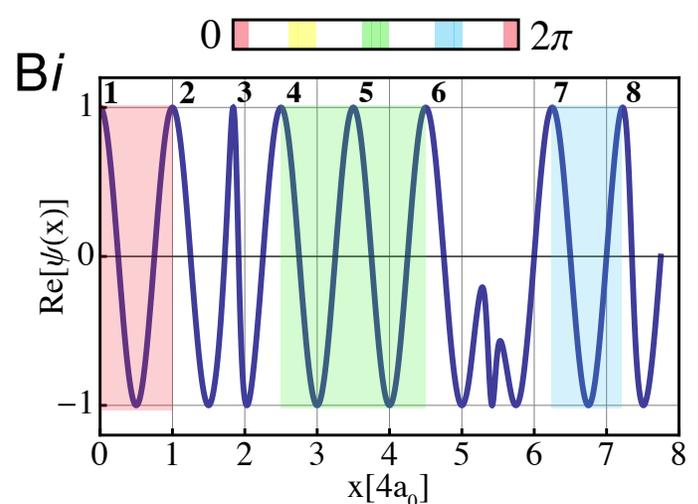
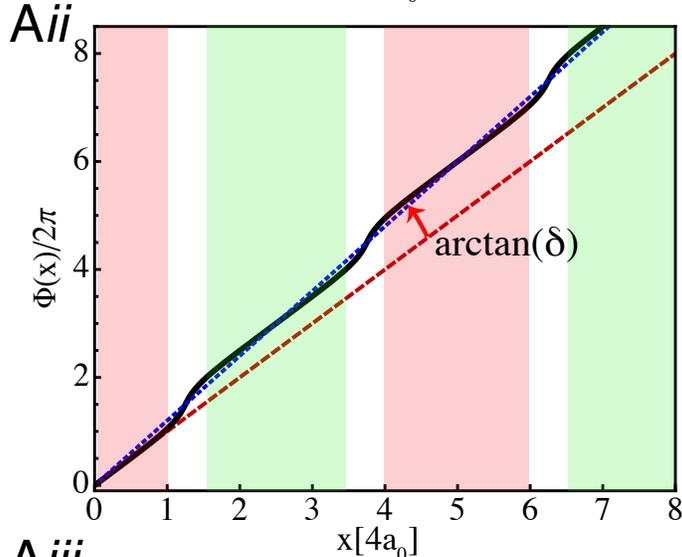
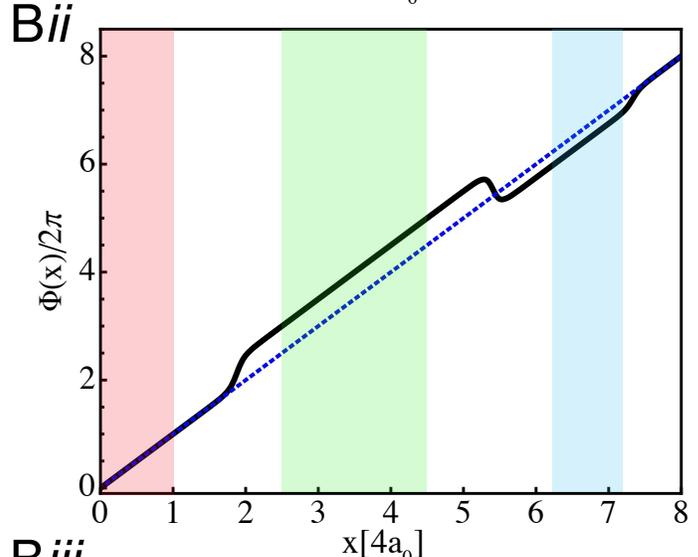
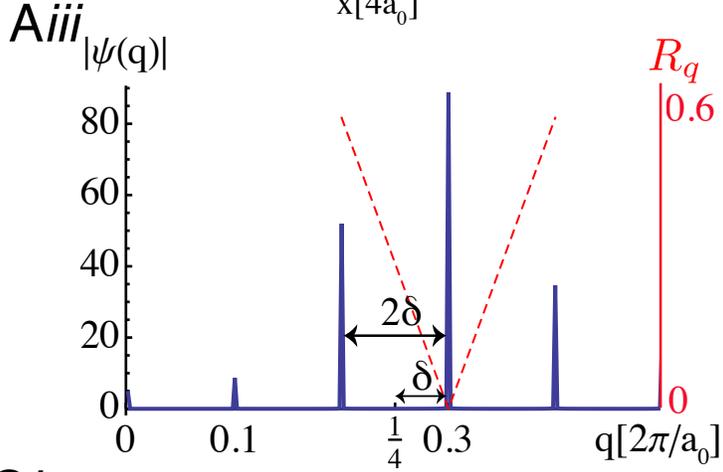
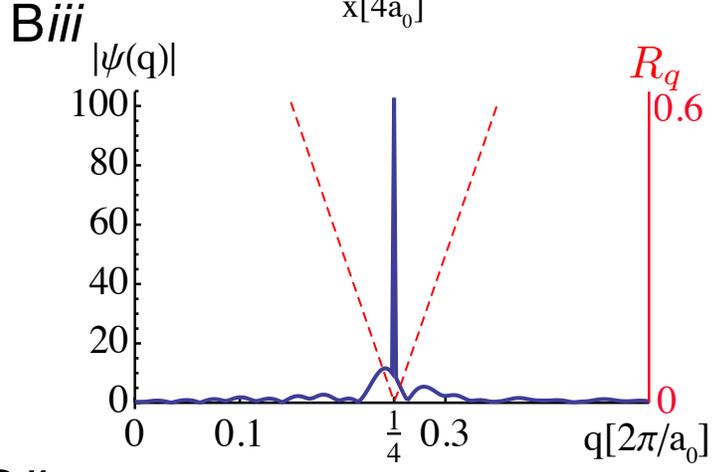
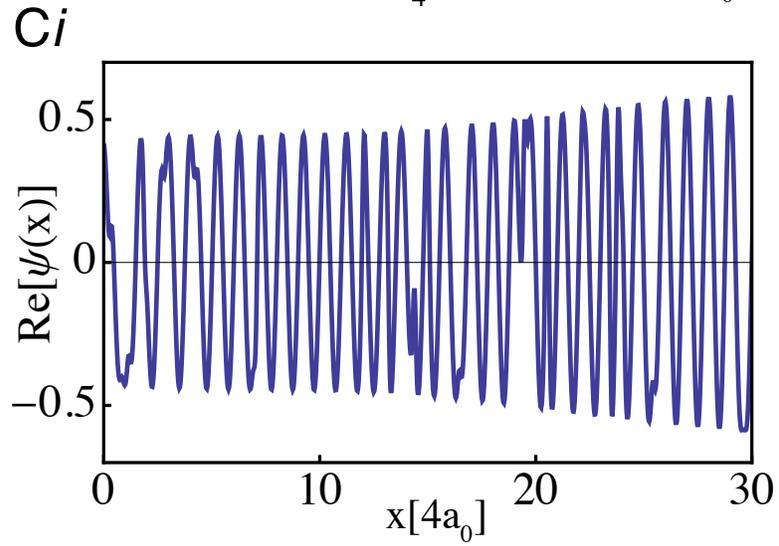
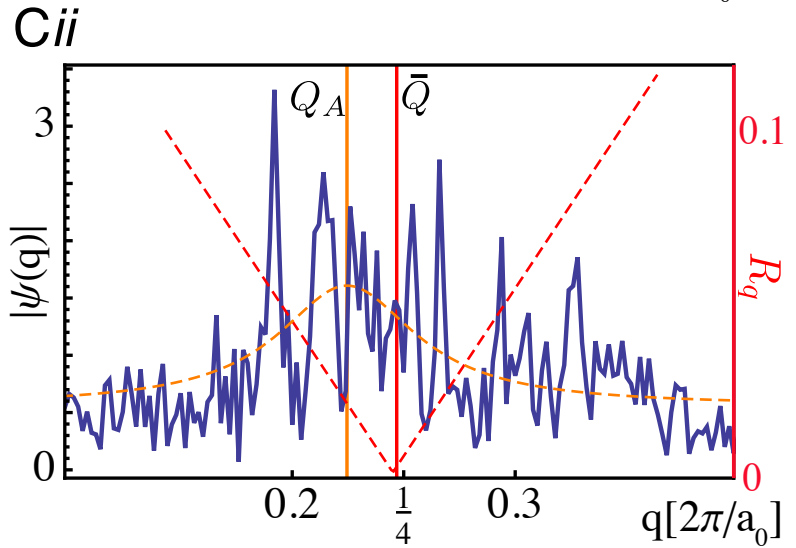

Figure 3

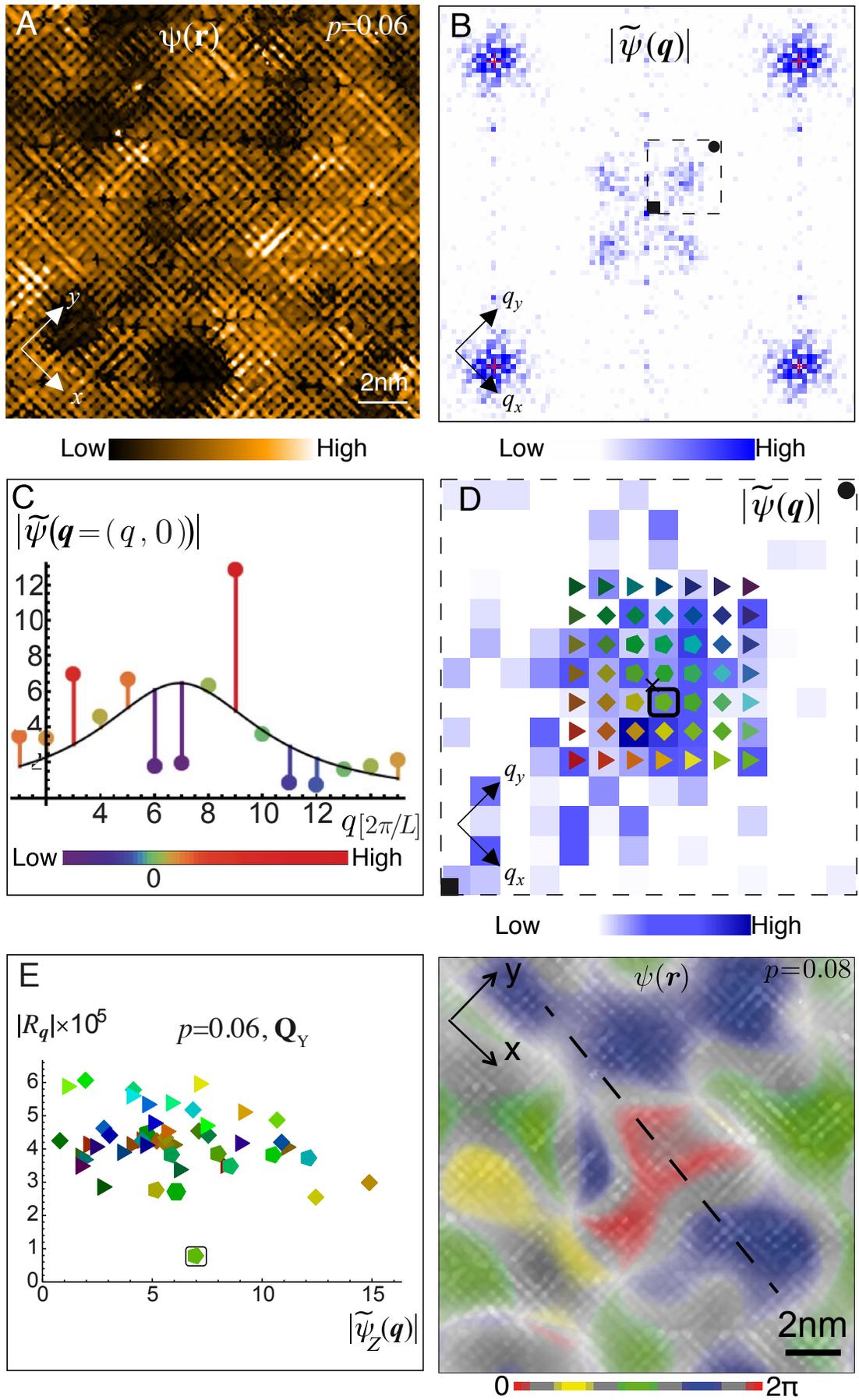

Figure 4

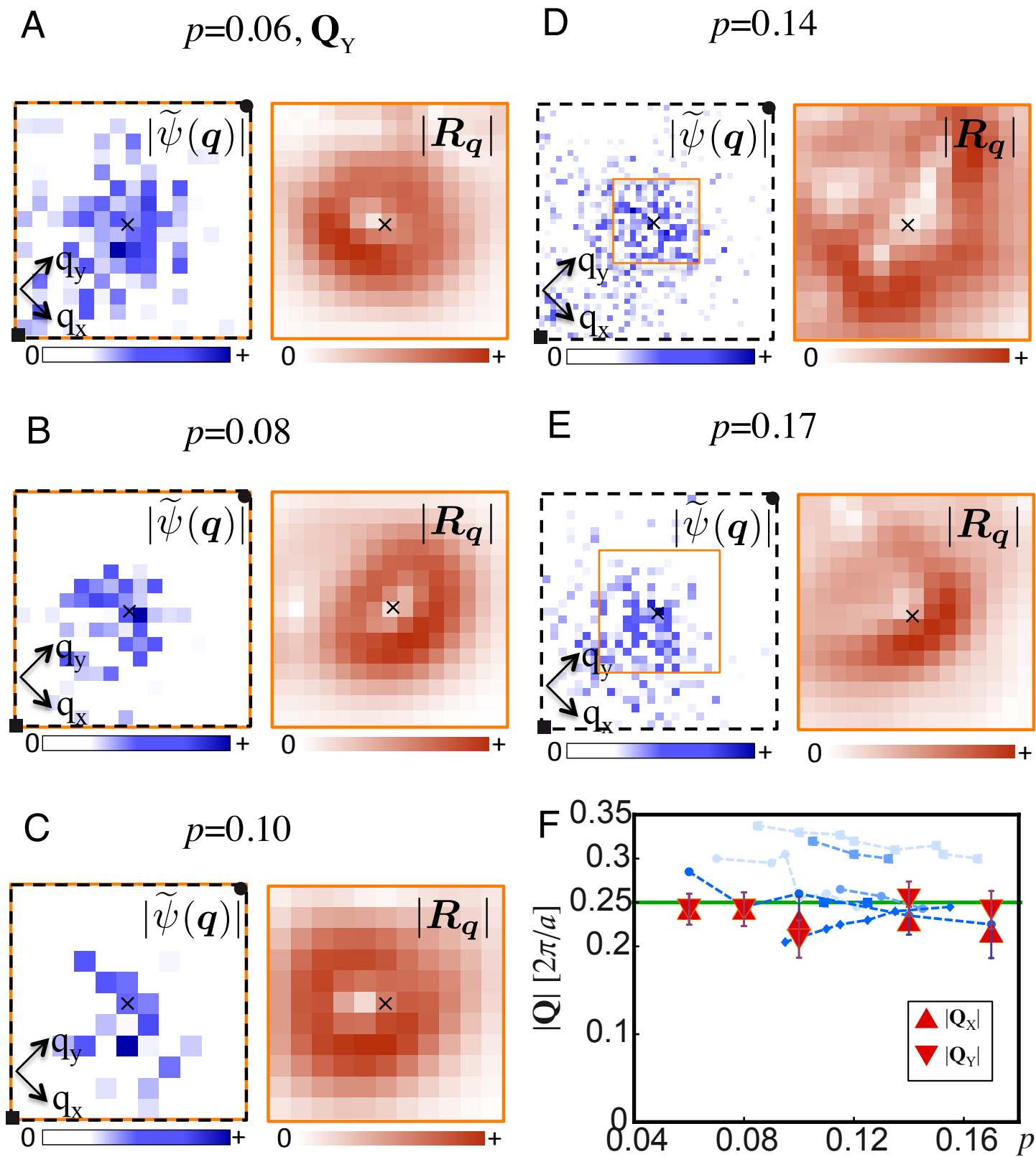

**Supporting Information**

# Commensurate $4a_0$-period Charge Density Modulations throughout the $Bi_2Sr_2CaCu_2O_{8+x}$ Pseudogap Regime

A. Mesaros, K. Fujita, S.D. Edkins, M.H. Hamidian, H. Eisaki, S. Uchida, J.C. Séamus Davis, M. J. Lawler and Eun-Ah Kim

**SI Text**

**1. *d*-symmetry form factor of CDM**

For a given BSCCO sample SI-STM measurement of the *Z*-map, where $Z(\mathbf{r}, E)=g(\mathbf{r}, +E)/g(\mathbf{r}, -E)$, we calculate the *d* symmetry form factor (dFF) $\tilde{\psi}(\mathbf{q})$ directly in Fourier space, which is equivalent to the procedure introduced in ref. 17 as,

$$\tilde{\psi}(\mathbf{q}) = \sum_{\mathbf{G}} \tilde{Z}(\mathbf{q} - \mathbf{G}) f(\mathbf{G}) \quad [S1.1]$$

where $\mathbf{G}$ are reciprocal lattice vectors of Z-map Fourier transform $\tilde{Z}(\mathbf{q})$, and $f(\mathbf{q})$ is the Fourier transform of an arbitrary smooth function on the sublattice scale which is used to sample the *Z*-map data at oxygen sites. We choose $f(\mathbf{G})$ to be 1 for the nine $\mathbf{G}$'s closest to origin, and zero for others.

**2. Heterogeneity and demodulation residue**

The CDM observed in cuprates are short-ranged, and disorder in the modulations challenge the measurement of their orientation, symmetry and wavevector. In Fig. S1 we demonstrate the effect of disorder on two CDM scenarios from Fig. 2 of main text. To generate discommensurations themselves in our simulations, e.g., in Fig. 2 in the main text, we rely on a Ginzburg Landau description of CDM detailed in *SI Text* section 5.

In the case of DC lattice (Fig. S1*A*) we consider variation in sizes of commensurate domains, keeping the incommensurability fixed (Fig. S1*B*), and we add smooth variation of CDM amplitude (Fig. S1*A*). The initially singular Fourier amplitude satellite peaks (Fig. 2*Aiii* in the main text) can completely disappear in the generated noise (Fig. S1*C*). Observation of DC lattice by amplitude-measuring scattering probes can therefore be hindered by disorder, as is demonstrably the case in layered dichalcogenides. In the case of randomly positioned discommensurations having random phase-slips (Fig. 2*Bi* in the main text and Fig. S1*E*), we add disorder in amplitude using smooth variation (Fig. S1*D*). The resulting Fourier amplitudes (Fig. S1*F*) show a broad irregular peak with high values at multiple wave vectors. This situation is characteristic of BSCCO SI-STM data, and can be generated in the case of periodic DC array (Fig. S1*B*), e.g., by adding random DC with random phase-slips. To avoid ambiguities in determining the average wave vector from Fourier amplitudes (see below and *SI Text* section 3), we introduce a phase-sensitive measurement based on demodulation residue, which robustly determines the average wavevector with high precision (Fig. 2*Cii* in the main text and Figs. S1*F*).

To describe the phase-sensitive measurement of average wave vector, we consider a generally short-ranged one-dimensional smooth modulation $\psi(x)$, defined to spatially average to zero. The case in Fig. 2*Bii* in the main text is an example representing BSCCO, however the following analysis is relevant for any disordered $\psi(x)$, not necessarily a case of commensurate wave dominated by discommensurations. The demodulation $\psi_q(x) = exp[-i\,q\,x]\,\psi(x)$ has the wavevector $\bar{Q} - q$, where $\bar{Q}$ is the wavevector of $\psi(x)$. The $\psi_q(x)$ has a phase that is, by definition, a fluctuating function of $x$ with zero average value.

Generally, any measurement samples the modulation in some finite region $0 < x < L$, so that the randomly fluctuating phase might accidentally have a small average value, i.e., have component linear in $x$; this effect however rapidly decays with $L$ for typical Gaussian disorder.

According to Eq. 1 of main text, we can interpret $\psi_q(x)$ as having zero wavevector while its phase argument $\Phi_q(x)$ fluctuates around a linear in $x$ term that is proportional to $\bar{Q} - q$. The value $q = \bar{Q}$ corresponds to vanishing linear term,

$$S_{q=\bar{Q}} \equiv \int_0^L \frac{dx}{L} \partial_x \left(\Phi_{q=\bar{Q}}(x)\right) \equiv 0. \tag{S2.1}$$

The $S_q$ tracks the accumulation of phase error in the (smooth) periodic function such as the $\Phi_q(x)$ might be. The phase argument function itself cannot be unambiguously extracted from data, suggesting we replace it with the well-defined periodic factor $exp[i\,\Phi_q(x)]$, which gives the equivalent $\int_0^L \frac{dx}{L} exp[-i\,\Phi_q(x)] \partial_x(exp[i\,\Phi_q(x)]) \equiv 0$. However, the derivative of the exponential factor can be singular due to sharp phase-slips. The physically meaningful expression must use the full, smooth, complex function $\psi_q(x)$ which gives the equation

$$R_{q=Q} \equiv \int_0^L \frac{dx}{L} |\psi_{q=Q}(x)|^2 \partial_x \Phi_{q=Q}(x) \equiv 0, \tag{S2.2}$$

i.e., the demodulation residue figure of merit introduced in the main text. The demodulated pattern with the phase-optimized choice $Q$ therefore has the minimal amount of spatial modulation. In an example of uniform modulation (long-range CDW with no disorder), the optimally demodulated function is strictly a constant equal to the amplitude of the modulation, and $Q = \bar{Q}$.

In presence of disorder, phase-optimized $Q$ rigorously equals $\bar{Q}$ in either of two limits: (i) When spatial fluctuations of complex amplitude $|\psi_Q(x)|$ are negligible compared to its spatial average value; (ii) When the complex phase spatially fluctuates on shorter scales than the complex amplitude does, even if the amplitude fluctuates strongly. The density waves observed in SI-STM images of BSCCO have primarily phase disorder, putting them close to the first limit.

We emphasize that even if the randomly disordered modulation strongly violates conditions for both limits (i) and (ii), the error $|Q - \bar{Q}|$ is still bounded by the characteristic wavevector, $k$, of amplitude fluctuations. Namely, consider $R_{Q+q}$: As $q$ grows the complex phase of demodulation varies faster in space (due to the $q \cdot x$ term), so that at values $|q| > |k|$ the condition of limit (ii) is satisfied, making $R_{Q+q} \approx S_{Q+q}$. Therefore, the functions $S$ and $R$ can deviate from each other only within a region of width $k$, hence

$$|Q - \bar{Q}| \lesssim k. \qquad [S2.3]$$

Note that by its definition $S_{\bar{Q}+q} \approx q$ through a large portion of Brillouin zone, and it is strictly periodic in $q$. The linear behavior is also observed in demodulation residue $R$, see Figs. 2 and S1.

To determine $\bar{Q}$ of short-ranged modulations, conventionally one seeks the maximum position, $Q_A$, of a fit to the broad and irregular peak of Fourier amplitudes, although the fit can be very poor (*SI Text* section 3). The use of phase-optimized $Q$ is especially necessary when the amplitude peak is asymmetric in which case $Q_A$ deviates significantly from $\bar{Q}$, e.g., in Fig. 2*Cii*. With a roughly symmetric peak the $Q_A$ might be close to $Q$ and give a good estimate of $\bar{Q}$, e.g., in Fig. S1*F*. One way to generate the former case with large mismatch of $Q_A$ and $\bar{Q}$ is by having long-range fluctuations in phase-slips, as illustrated by the discommensuration arrangement in Fig. S1*G*, used to obtain Fig. 2*Cii* in the main text, compared to case in Fig. S1*E*, used to obtain Fig. S1*F*. In BSCCO SI-STM data, the mismatch of commensurate $Q$ and doping-dependent (ref. 33) $Q_A$ is largest at lowest doping, and we suspect this is caused by propensity for smaller random commensurate domains at lower doping, as illustrated in Fig. S1*H*.

Additionally, in two dimensions there are topological point-like defects, dislocations (Fig. S2*A*), where the considered demodulation $\psi_q(x)$ has a vortex in its phase argument. These are especially relevant for SI-STM data on BSCCO. A singularity of the phase argument function can only occur at the point where the complex amplitude $|\psi_q(x)|$ vanishes. The density of two-dimensional demodulation residue $R_q^\alpha(x)$ (see *SI Text* section 4) is therefore smooth at the position of dislocation, even though it contains derivatives. Crucially, the local disturbance in $R_q^\alpha(x)$ created by a dislocation spatially averages to zero, see Figs. S2*B*, *C*. The dislocations therefore generally do not influence the demodulation measurement of the phase-optimized wavevector.

### 3. Statistical analysis: two-dimensional fitting

Given the dFF data, we focus on a single characteristic intensity peak (out of two, $X$ or $Y$, centered on the $x$ and $y$ axis, respectively) by restricting $q$ to a square covering a quarter of the first Brillouin zone (e.g., dashed square in Fig. 1*D* in the main text). This restricted data is simply referred to as "dFF data" and labeled $\widetilde{\Psi}(q)$ within this section.

The Fourier amplitude of dFF, $|\tilde{\psi}(q)|$, typically shows a broad intensity distribution with peaks at multiple $q$ values, as presented for all samples in Fig. S3. The standard procedure for describing the modulation in this situation involves a two-dimensional fit of $|\tilde{\psi}(q)|$ with a single-peaked smooth function, reporting the position of that peak. We apply the two-dimensional fit of several smooth functions: the offset Gaussian distribution $c + a\, exp[-(q-q_0)^2/2s^2]$, offset Lorentz distribution $c + a/[(q-q_0)^2 + s^2]$, two-variate Cauchy distribution $c + a/[(q-q_0)^2 + s^2]^{3/2}$, and their versions with different widths $s$ in orthogonal directions. For example, the data presented in Fig. S3*A* is best fitted by the shifted Lorentzian centered at $q_0 = (0.21, 0.22) \times \frac{2\pi}{a}$, with width $s = 0.11 \times \frac{2\pi}{a}$.

We however observe that the fit is quite poor, with values of adjusted R-squared of about 0.7, which corroborates visual inspection, e.g., in Fig. 3*C of main text*. The culprit for the poor fit is data having multiple local (single pixel) peaks in an asymmetric distribution. We confirm this by showing that statistically these peaks cannot be described by the simple smooth single-peaked distribution.

A standard tool for analyzing the nature of fit is the normal Quantile-Quantile (QQ) plot, shown for all the samples in Fig. S4. The fit residuals, i.e. the set of differences between a value $|\tilde{\psi}(\boldsymbol{q})|$ and its fitted value, are sorted by size and indexed by $n = 1, \ldots, N$. It is natural to expect that data points are randomly scattered around the fit function, producing a normal distribution of fit residuals. In the QQ plot, the residuals are plotted against the inverse cumulative distribution of the standardized normal distribution evaluated at $(n - 3/8)/(N + 1/4)$ (ref. 45). If the fit residuals are indeed normally distributed the QQ plot follows a straight line centered at the origin. We fit the part of QQ plot near the origin to a straight line (black line in Fig. S4). However, for a typical dataset in Fig. S4 the significant deviations from the straight line at the high values of residuals are visually striking. They show that the largest values of $|\tilde{\psi}(\boldsymbol{q})|$ deviate from a normal scatter around a single smooth peaked fit function.

This deviation is statistically significant, according to the comparison of the highest value of fit residual with the probability distribution for the highest value drawn from a normal distribution, this normal distribution best describing the fit residuals. The observed highest residual value typically lies within 3 to 6 standard deviations away from the center of this probability distribution.

**4. Demodulation in two dimensions, smoothing, and optimization error**

To apply our analysis to two-dimensional data, the SI-STM $Z$-map dFF $\psi(\boldsymbol{x})$, we define the demodulation directly in Fourier space as

$$\Psi_{\boldsymbol{q}}(\boldsymbol{k}) = exp\left(-\frac{(\boldsymbol{k} - \boldsymbol{q})^2}{2\Lambda^2}\right) exp(-i\, \boldsymbol{q} \cdot \boldsymbol{x})\, \tilde{\psi}(\boldsymbol{q} + \boldsymbol{k}) \qquad [S4.1]$$

or in words, we shift the Fourier space origin to $\boldsymbol{q}$ and filter by a Gaussian suppression around it, which is equivalent to a real-space Gaussian smoothing of $exp(-i\,\boldsymbol{q}\cdot\boldsymbol{x})\,\psi(\boldsymbol{x})$ on lengthscale $1/\Lambda$ (Fig. S5*A*) (ref. 46). The analysis of CM requires smoothing because one needs to include only the data within a single broad intensity distribution (either the one around the $x$ or $y$ axis) from the Fourier data of $\psi(\boldsymbol{x})$. Our choice of smoothing procedure, centered on $\boldsymbol{q}$, instead of smoothing the whole broad distribution once, gives equal regard to every demodulation wavevector $\boldsymbol{q}$ (Fig. S5*A*). Note that the demodulated pattern $\Psi_{\boldsymbol{q}}(\boldsymbol{x})$ is complex since data around $-\boldsymbol{q}$ is not used (Fig. S5*B*).

The demodulation residue in two dimensions is a vector with components $\alpha = x, y$, (Figs. S5*C*, *D*) and in field-of-view of finite size $L$ is a direct generalization of one dimensional case:

$$\boldsymbol{R}_{\boldsymbol{q}}^{\alpha}[\psi] \equiv \int \frac{d^2\boldsymbol{x}}{L^2} Re\big[\Psi_{\boldsymbol{q}}^{*}(\boldsymbol{x})(-i\partial_\alpha)\Psi_{\boldsymbol{q}}(\boldsymbol{x})\big]. \qquad [\text{S4.2}]$$

We find that the demodulation residue is robust to choice of value of filtering cutoff $\Lambda$ in all underdoped samples. The phase-optimized wavevector chosen by demodulation residue varies negligibly as long as $\Lambda$ is larger than the spacing between $\boldsymbol{q}$-points (the $\boldsymbol{q}$-space is discrete with spacing $\frac{2\pi}{L}$) and a few times smaller than the distance between neighboring intensity peaks in Fourier data of $\psi(\boldsymbol{x})$.

As discussed in *SI Text* section 2, the phase-optimized wavevector gives a linear fit $\boldsymbol{Q}\cdot\boldsymbol{x}+\varphi$ to the phase argument function $\Phi(\boldsymbol{x})$ of the modulation $\psi(\boldsymbol{x})$ (see Fig. 2*B*, *E* in the main text for one-dimensional examples). To estimate the error of the obtained slope $\boldsymbol{Q}$, we calculate:

$$\sigma^2 \equiv \frac{\int \frac{d^2\boldsymbol{x}}{L^2} v^2}{\int \frac{d^2\boldsymbol{x}}{L^2} |\Psi_{\boldsymbol{Q}}(\boldsymbol{x})|^2}, \qquad [\text{S4.3}]$$

where $v_\alpha = Re\big[\Psi_{\boldsymbol{Q}}^{*}(\boldsymbol{x})(-i\partial_\alpha)\Psi_{\boldsymbol{Q}}(\boldsymbol{x})\big]$. The $\sigma$ corresponds to the standard deviation (caused by spatial fluctuations) of the phase argument function around the linear term $\boldsymbol{Q} \cdot \boldsymbol{x} + \varphi$ (hence it is evaluated exactly at the phase-optimized wavevector $\boldsymbol{Q}$). In the one-dimensional examples of Figs 2*Aii*, 2*Bii* in the main text the $\sigma$ quantifies the standard deviation of the phase function around the optimal fit (dotted blue line). The $\sigma$ has units of wavevector due to the use of normalized demodulation function $\Psi_Q(x)$, and directly quantifies the error in measurement of phase optimized $\boldsymbol{Q}$. For a discrete dataset, such as an SI-STM image, we replace: the derivative by difference at neighboring pixels, the integral by sum over all image pixels, the $L^2$ by total number of pixels; and finally the wavevector error $\sigma$ is obtained (and reported in Fig. 4*F* in the main text) by simple conversion of units from $1/pix$ to $2\pi/a_0$, where $pix$ is the real-space width of an image pixel.

## 5. CDM commensurability in underdoped BSCCO

When we extract the phase-optimized wave vector, representing the average wave vector of CDM, we find values consistent with commensurate $Q_{phase-opt} = Q_0 = \frac{1}{4}\frac{2\pi}{a}$ throughout underdoped regime of BSCCO (see *SI Text* section 6). We now demonstrate the consistency of this finding and simultaneously show that the local wavevector is also $Q_0$. The CDM here have non-overlapping uniaxial nature, so we focus on one direction, e.g., $\boldsymbol{q} = q\boldsymbol{e}_x$ in Fig. S6. The CDM order parameter at $q = Q_0$, Fig. S6*A (see also Fig 3F and S1H)*, shows multiple domains with CDM phase close to the commensurately pinned values $n\frac{2\pi}{4}$ (see end of this Section), indicating the local commensurate wave vector $Q_0$. In real space SI-STM image (Fig. S6*B*), areas where CDM order parameter amplitude is high consistently show commensurate patches of typical $4a_0$-wide dFF pattern (Fig. 1*B*, *F* in the main text) and reveal these patches forming commensurate domains. Every patch individually is therefore aligned with underlying lattice, an effect observed already in ref. (47), but the patches are shifted in position with respect to each other by values $a_0$, $2a_0$ or $3a_0$.

As in our earlier analysis (ref. 18) we here observe that CDM order parameter phase fluctuations are dominated by phase vortices (white circle in Fig. S6*A*), i.e. CDM dislocations, which seems to indicate an incommensurate CDM. In commensurate CDM, dislocations also exist as points where multiple discommensurations meet and provide a total of $2\pi$ phase slip upon encircling the meeting point, and we observe this directly in real-space. In vicinity of order parameter vortex, in real space we find a few patches of appreciable OP amplitude representing commensurate domains meeting at the defect. Since here the order parameter's smoothing length scale (see *SI Text* section 4) approaches the commensurate domain size, the discommensuration meeting point smooths into a phase vortex in Fig. S6*A*.

Although our data shows local wave vector $Q_0$ and phase slips of various multiples of $\pi/2$ between multiple domains (Fig 3F, Fig. S1H and Fig. S6), in principle there could be an underlying DC lattice and an incommensurate average wave vector observed over distances larger than SI-STM field-of-view. The error in phase-optimized wave vector $Q_{phase-opt}$ induced by adding/removing into field-of-view a single DC of such a sparse underlying DC lattice is consistent with error bar for $Q_{phase-opt}$ obtained in *SI Text* section 4. We note that by definition of phase-optimized wavevector, an ideal DC lattice would appear as a see-saw in phase of CDM order parameter, e.g., see deviation of phase argument from best-fit line in Fig. 2Aii in the main text. We do not see such features in SI-STM images.

To generate discommensurations (DC's) in our simulations (Figs. 2 in the main text, Figs. S1, S8), we use a Ginzburg Landau order parameter theory for CDM, which in cuprates has square lattice symmetry (ref. 48). The theory in ref. 48 deals with the commensurate CDM state in a region around doping 1/8, but does not deal with DC's. Compared to well-known CDW systems in layered dichalcogenides (ref. 49, 50), the energy term describing the local commensuration tendency appears as a quartic term instead of as a cubic. We focus on the CDM phase mode, and find an ideal DC lattice as a possible ground state (ref. 19). The quartic term prefers $Q_0 = \frac{1}{4}\frac{2\pi}{a}$ commensurate CDM with phase locked at multiples of $\frac{2\pi}{4}$. The ideal DC lattice with phase slips of $n\frac{2\pi}{4}$, $n$ integer, and separation distance $n/(4\delta)$, describes an array of commensurate domains that accumulates CDM phase and gives average wave vector $\bar{Q} = Q_0 + \delta$ when averaged over multiple domains. The Fourier satellite peaks of DC lattice are separated by $4\delta/n$ around $\bar{Q}$, while $Q_0$ amplitude vanishes (Fig. 2B shows example of $n = 2$). The spatial profile of DC lattice we obtained as a variational solution (ref. 19) using 20 harmonics $sin\ (j\ 4\ \delta\ x), j = 1\ ...\ 20$.

## 6. Summary of results for all samples

Fig. S7 shows the demodulation residue for modulations along *x*-axis, calculated for all analyzed samples (for *y*-axis modulations see Fig. 4 in the main text). For underdoped samples a single phase-optimized wavevector is apparent. For higher dopings the distinction is not so clear and the optimum is within a larger area of Fourier space, in accord with the weakening of the dFF component at those dopings. The results are robust to changes of filtering cutoff $\Lambda$ for the underdoped samples (see *SI Text* section 4).

## 7. Global CDM lattice commensurability

We introduce the following hypothesis: Differences in measured CDM wave vectors over different cuprate families are due to different arrangements of discommensurations (DC's) in the underlying $4a_0$ periodic order. In support, we now relate cuprate materials to our simulated DC arrangements in Figs. 2 in the main text, Fig. S1 and S8, to show consistency with various values of extracted wavevectors and consistency with varying widths of the CDM peaks. A clear signature of our proposal would be observation of satellite Fourier peaks in scattering probes when DC lattice is present in a material, however, we have shown how the satellites might be unobservable due to disorder (e.g., compare Fig. 2*Aiii* in the main text and Fig. S1*C*), and this effect is already recognized in layered dichalcogenides (ref. 34).

In BSCCO, our finding of $4a_0$ commensurate patches and commensurate average wavevector (*SI Text* section 5) are simulated in a one-dimensional cut by a random distribution of commensurate domains of typical size around $8a_0$, with discommensurations whose phase-slips are random multiples of $2\pi/4$ (Figs. S1*D*, *E*, *G* and 2*Bi*, 2*Bii* in the main text). Irregular Fourier amplitude peak with phase-optimized wavevector $\bar{Q} = \frac{1}{4}\frac{2\pi}{a_0}$ mirrors our BSCCO findings. Our simulated one-dimensional CDMs (Fig. 2*Cii* in the main text or Fig. S1*F*) have similar full-width $\Delta Q \sim 0.05 \frac{2\pi}{a_0}$ of intensity peak as the BSCCO STS data (Fig. 3*B* in the main text).

In comparing to YBCO, existing X-ray measurements (ref. 15) offer two restrictions: (i) The CDM intensity peak is centered on wavevector $\bar{Q} \sim 0.3 \frac{2\pi}{a_0}$; (ii) The intensity peak has full-width $\Delta Q \sim 0.02 \frac{2\pi}{a_0}$. These do not rule out a commensurate local wavevector $Q_0 = 1/4$: Fig. 2*Aiii* in the main text shows how in absence of disorder a DC lattice generates a sharp Fourier peak at $\bar{Q} = 0.3 \frac{2\pi}{a_0}$. The DC lattice also generates strong satellite peaks which are not yet reported in YBCO. Adding disorder can suppress the satellites, while giving width $\Delta Q \sim 0.02 \frac{2\pi}{a_0}$ to the intensity peak, see Fig. S1*C*.

Neutron scattering in LNSCO (see ref. 38 and references within) shows $\bar{Q}$ near $0.23 \frac{2\pi}{a_0}$, therefore a small incommensurability $\delta = -0.02$, and sharp peaks with full-width $\Delta Q \sim 0.01 \frac{2\pi}{a_0}$. Within our hypothesis this situation reflects a DC lattice as demonstrated in Fig. S8. A total of $2\pi$ phase slip is expected on lengthscale of over 20nm. Note that the dominant satellite peak at $0.31 \frac{2\pi}{a_0}$ (i.e., at distance $4\delta$ from $\bar{Q}$ since our chosen DC phase slips are $-\pi/2$) is suppressed by randomness in DC positions within their lattice. Our proposal is an alternative to the existing model of CDM patches having fixed but different widths (ref. 38).

X-ray scattering on LBCO (ref. 14) finds a peak near $\bar{Q} = 0.23 \frac{2\pi}{a_0}$, with small incommensurability $\delta = -0.02$ compared to $Q_0 = \frac{1}{4}\frac{2\pi}{a_0}$, and having a larger $\Delta Q \sim 0.04 \frac{2\pi}{a_0}$. The increased peak width compared to above LNSCO scenario can simply be caused by any of several factors such as increased CDM amplitude variation, more randomness in positions of DC within their lattice, or additional random DCs which average to zero phase-slip and occur on top of the underlying DC lattice.

We uncovered the problem of irregular Fourier amplitude distribution within a broad CDM intensity peak and it is presently unclear if this property can be observed in other cuprates. In the BSCO family REXS experiments (refs. 42, 43) have a high wavevector resolution and clearly show these broad irregular Fourier amplitude distributions. On the other hand, X-ray experiments on YBCO show smoother Fourier intensity peaks, but one must focus on the wavevector resolution: typically (refs. 44, 51) there are only about 5 distinct amplitude values in a one-dimensional cut of $q$-space through the intensity peak, which washes out the possible irregularity. In comparison, STS data used here has of order 50 amplitude values in a two-dimensional profile of the intensity peak (Fig. 3D in the main text).

**Fig. S1.**

Discommensuration model in panels A, B, C is for a situation which may apply to YBCO. Panels D, E, F and G are models of our findings in BSCCO.

(A) Modulation is the real part of complex wave $\psi(x) = A(x)e^{i(Q_0 x + \varphi(x))}$ having commensurate domains with local wave vector $Q_0 = \frac{1}{4} \times \frac{2\pi}{a}$ (period $4a_0$). The amplitude $A(x) \geq 0$ varies smoothly around value 1, seen as the envelope of modulation. Phase slips are incorporated in $\varphi(x)$ (see B). The average $\bar{Q} = 0.3 \times \frac{2\pi}{a}$.

(B) The local phase $\varphi(x)$ of $\psi(x)$ in A, constructed as a discommensuration (DC) array in the phase argument $\Phi(x) = Q_0 x + \varphi(x)$. Phase slips of all DC's are set to $+\pi$. The distances between neighboring DC's vary randomly around average distance set by value of incommensurability $\delta = \bar{Q} - Q_0 = 0.05 \times \frac{2\pi}{a}$ (*SI Text* section 5).

(C) Fourier amplitudes $|\tilde{\psi}(q)|$ of the modulation $\psi(x)$ in A (blue line) show narrow peak at $\bar{Q} = 0.3 \times \frac{2\pi}{a}$. The satellites depend on spatial profile of DC's, but here they are strongly suppressed by disorder. The calculated phase-sensitive figure of merit, demodulation residue $|R_q|$ (red dashed line), as a function of $q$ has the minimum exactly at the average $\bar{Q}$.

(D) Modulation is the real part of complex wave $\psi(x) = A(x)e^{i(Q_0 x + \varphi(x))}$ having commensurate domains with local wave vector $Q_0 = \frac{1}{4} \times \frac{2\pi}{a}$ (period $4a_0$). The amplitude $A(x) \geq 0$ varies smoothly around value 1, seen as the envelope of modulation. Phase slips are incorporated in $\varphi(x)$ (see E).

(E) The local phase $\varphi(x)$ of $\psi(x)$ in D, constructed as a discommensuration (DC) array in the phase argument $\Phi(x) = Q_0 x + \varphi(x)$. Phase slips of DC's are random integer multiples of $\frac{\pi}{2}$, and they all cancel across x-axis to retain the average $\bar{Q} = Q_0$. Distances between neighboring DC's vary randomly. The spatial profile of each DC is obtained as variational solution of Ginzburg Landau theory for CDM phase, using 20 harmonics for $\delta = 0.01$ (*SI Text* section 5).

(F) Fourier amplitudes $|\tilde{\psi}(q)|$ of the modulation $\psi(x)$ in $D$ (blue line) have a broad irregular intensity peak at $\bar{Q} = Q_0$. The calculated phase-sensitive figure of merit, demodulation residue $|R_q|$ (red dashed line), as a function of $q$ has the minimum exactly at the average $\bar{Q}$.

(G) The local phase $\varphi(x)$ used to create Fig. 2Ci, 2Cii in the main text, constructed as a random discommensuration (DC) array in the phase argument $\Phi(x) = Q_0 x + \varphi(x)$. The array is generated in the same way as in $E$, except that it has double density of DCs, and the random distribution of phase-slip signs has a long-range correlation.

(H) The complex phase of d-form factor order parameter $\Psi_{\bar{Q}}(x)$ for CDM along x-axis, obtained at dopings 0.06 (left panel) and 0.14 (right panel). The demodulation is by phase-averaged $\bar{Q}$ of length $\frac{1}{4}\frac{2\pi}{a}$ in both cases. The complex phase of $\Psi_{\bar{Q}}(x)$ is color-coded (legend below), while complex amplitude varies the color saturation from white (zero amplitude) to full (maximal amplitude in image). Both panels show a field-of-view of approximately 32x32 unit-cells, and smoothing parameter $\Lambda$ is the same. Commensurate domains seem to be more disordered at lower dopings.

**Fig. S2.**

(A) A spatial modulation in two dimensions, with a dislocation created in the center of image by a $2\pi$ vortex in the complex phase and a suppression of complex amplitude by a Gaussian of width one modulation period: $Re[\psi(x,y)]$ with $\psi(x,y) = \exp\left(-\frac{1}{2}(x^2 + y^2)\right)\exp\left(i\,Q\,x + i\,\text{atan}\left(\frac{y}{x}\right) + i\frac{\pi}{2}\right)$.

(B) Demodulation residue density along $x$-axis evaluated for modulation $\psi(x,y)$ in $A$ at wavevector $Q\hat{x}$, which is the phase-optimized wavevector. Dislocation position marked by cross.

(C) Demodulation residue density along $y$-axis evaluated for modulation $\psi(x,y)$ in $A$ at wavevector $Q\hat{x}$, which is the phase-optimized wavevector. Dislocation position marked by cross.

**Fig. S3.**

The Fourier amplitude of dFF, $|\tilde{\Psi}(q)|$, with wavevector $q$ restricted to a square area with corner at the Fourier space origin (black square) and center at $\boldsymbol{Q}_X = \frac{1}{4}\boldsymbol{G}_X$ or $\boldsymbol{Q}_Y = \frac{1}{4}\boldsymbol{G}_Y$, where $\boldsymbol{G}_X$ and $\boldsymbol{G}_Y$ are the Bragg peaks. This area covers the intensity distribution of a unidirectional CM. The color-coded quantity is the fit residual to best fitting smooth two-dimensional peaked function (see text and Fig. S4). Data from BSCCO samples at doping level $p$ and having superconducting transition temperature $T_c$ is shown: $p = 0.06, T_c = 20$K for $\boldsymbol{Q}_X$ in $A$; $p = 0.06, T_c = 20$K for $\boldsymbol{Q}_Y$ in $B$; $p = 0.08, T_c = 45$K for $\boldsymbol{Q}_X$ in $C$; $p = 0.08, T_c = 45$K for $\boldsymbol{Q}_Y$ in $D$; $p = 0.10, T_c = 65$K for $\boldsymbol{Q}_X$ in $E$; $p = 0.10, T_c = 65$K for $\boldsymbol{Q}_Y$ in $F$; $p = 0.14, T_c = 74$K for $\boldsymbol{Q}_X$ in $G$; $p = 0.14, T_c = 74$K for $\boldsymbol{Q}_Y$ in $H$; $p = 0.17, T_c = 89$K for $\boldsymbol{Q}_X$ in $I$; $p = 0.17, T_c = 89$K for $\boldsymbol{Q}_Y$ in $J$.

**Fig. S4.**

The standard QQ plot compares the inverse cumulative normal distribution (horizontal axis) to the inverse cumulative distribution of fit residuals (vertical axis) for the dFF Fourier amplitude data, $|\tilde{\Psi}(q)|$, in Fig. S3. The residuals for each dataset are obtained from the best fitting of data in Fig. S3 to several single-peaked smooth two-dimensional functions (see text). The color-coding for different data points $|\tilde{\Psi}(q)|$ is used in Fig. S3. Black line is linear fit to QQ plot obtained for half the data points nearest to origin. Data from BSCCO samples at doping level $p$ and having superconducting transition temperature $T_c$ is shown: $p = 0.06, T_c = 20K$ for $Q_X$ in A; $p = 0.06, T_c = 20K$ for $Q_Y$ in B; $p = 0.08, T_c = 45K$ for $Q_X$ in C; $p = 0.08, T_c = 45K$ for $Q_Y$ in D; $p = 0.10, T_c = 65K$ for $Q_X$ in E; $p = 0.10, T_c = 65K$ for $Q_Y$ in F; $p = 0.14, T_c = 74K$ for $Q_X$ in G; $p = 0.14, T_c = 74K$ for $Q_Y$ in H; $p = 0.17, T_c = 89K$ for $Q_X$ in I; $p = 0.17, T_c = 89K$ for $Q_Y$ in J.

**Fig. S5.**

(A) The dFF Fourier amplitude in $p=0.06$ sample, surrounding $Q_X$ within a square whose diagonal reaches halfway to Bragg peak. A demodulated complex pattern in real-space, $\Psi_q(x)$, shown in B, is defined by wavevector $q$ and is smoothed by Gaussian suppression of Fourier data beyond the cutoff $\Lambda$ (radius of orange circle).

(B) The complex phase of $\Psi_q(x)$ is color-coded, while complex amplitude varies the brightness of color from black (zero amplitude) to full brightness (maximal amplitude in image). Points around which the complex phase winds by $\pm 2\pi$ (green/purple) are dislocations.

(C) The density, along spatial direction $\alpha$, of demodulation residue at $q$ is defined as $R_q^\alpha(x) = Re[\psi_q^*(x)(-i\partial_\alpha)\psi_q(x)]$. The demodulation residue along the horizontal axis of image, $R_q^{hor}$, is the average of $R_q^{hor}(x)$, here multiplied by $10^5$.

(D) Same as C, along vertical direction (*ver*) instead of horizontal.

**Fig. S6.**

(A) The complex phase of $\Psi_{Q_{opt}}(x)$, the CDM order parameter (OP) at phase-optimized wavevector $Q_{opt} = \frac{1}{4}\frac{2\pi}{a}$ along $x$-axis, is color-coded, while complex amplitude varies the brightness of color from black (zero amplitude) to full brightness (maximal amplitude in image). A phase vortex is at the center of dashed circle.

(B) Real-space SI-STM Z-map for doping $p=0.08$, from which image A is created. The dashed circle corresponds to the one in A. Patches of width $4a_0$ (rectangles) are marked within regions of high OP amplitude identified in A, and their colors (as in A) represent dominant value of phase across the patch, matching well multiples of $2\pi/4$. Patches exhibit the typical dFF pattern (see *SI Text* section 5). The OP phase-vortex obtained by smoothing in A corresponds to mutual misalignment of the $4a_0$ patterns by multiples of $a_0$, indicating a dislocation at the meeting point of commensurate domains.

**Fig. S7.**

(A)-(E). For a series of samples with estimated hole densities $p=0.06, 0.08, 0.10, 0.14, 0.17$: Left panel shows measured amplitude $|\tilde{\psi}(q)|$, within the square region of $q$-space (bottom dashed square in Fig. 3B in the main text) bounded by $(0,0)$, $(\pi/2a, \pi/2a)$, $(\pi/a, 0)$, $(\pi/2a, -\pi/2a)$, which defines the CM orientation along $x$-axis; right panel shows the value of demodulation residue $|R_q| = \sqrt{(R_q^x)^2 + (R_q^y)^2}$ calculated, using cutoff $\Lambda = 0.08 \times \frac{2\pi}{a}$, within the square region of $q$-space marked by orange thin square on the left panel. The corresponding position of $(\frac{1}{4}, 0) \times \frac{2\pi}{a}$ is marked by a small cross. The pixel at which the $|R_q|$ is found to be a minimum is identified as the phase-optimized CM wavevector for that carrier density. The measured $|R_q|$ varies smoothly and drops quickly towards zero at a single discrete wavevector near center of image, similarly to behavior in one-dimension (e.g. Fig. 2 in the main text). The phase-optimized wavevector $\mathbf{Q}$ in continuous $\mathbf{q}$-space can be at most localized within the whitest pixel area, leading to a minimal error of $0.02 \times \frac{2\pi}{a}$. The error value $0.03 \times \frac{2\pi}{a}$ is obtained from spatial variation of residue function (see *SI Text* section 4).

**Fig. S8.**

Discommensuration model of a situation that may apply to LNSCO.

(A) Modulation is the real part of complex wave $\psi(x) = A(x)e^{i(Q_0 x + \varphi(x))}$ having commensurate domains with local wave vector $Q_0 = \frac{1}{4} \times \frac{2\pi}{a}$ (period $4a_0$). The amplitude $A(x) \geq 0$ varies smoothly around value 1, seen as the envelope of modulation. Phase slips are incorporated in $\varphi(x)$ (see B). The average $\bar{Q} = 0.23 \times \frac{2\pi}{a}$.

(B) The local phase $\varphi(x)$ of $\psi(x)$ in A, constructed as a discommensuration (DC) array in the phase argument $\Phi(x) = Q_0 x + \varphi(x)$. Phase slips of DC's are set to $-\pi/2$. The distances between neighboring DC's vary randomly around average distance set by value of incommensurability $\delta = \bar{Q} - Q_0 = -0.02 \times \frac{2\pi}{a}$ (*SI Text* section 5).

(C) Fourier amplitudes $|\tilde{\psi}(q)|$ (blue line) of the modulation $\psi(x)$ in A. Amplitudes are smoothed by Gaussian of width $1.4 \cdot 10^{-3} \frac{2\pi}{a} = 2 \cdot 10^{-3} \text{Å}^{-1}$. Narrow peak at $\bar{Q} = 0.23 \times \frac{2\pi}{a}$ is captured by both a Lorentzian fit to Fourier amplitude (dashed orange) and by phase-optimized wavevector at minimum of demodulation residue $|R_q|$ (dashed red line). Fourier satellites are strongly suppressed by the randomness in DC array.

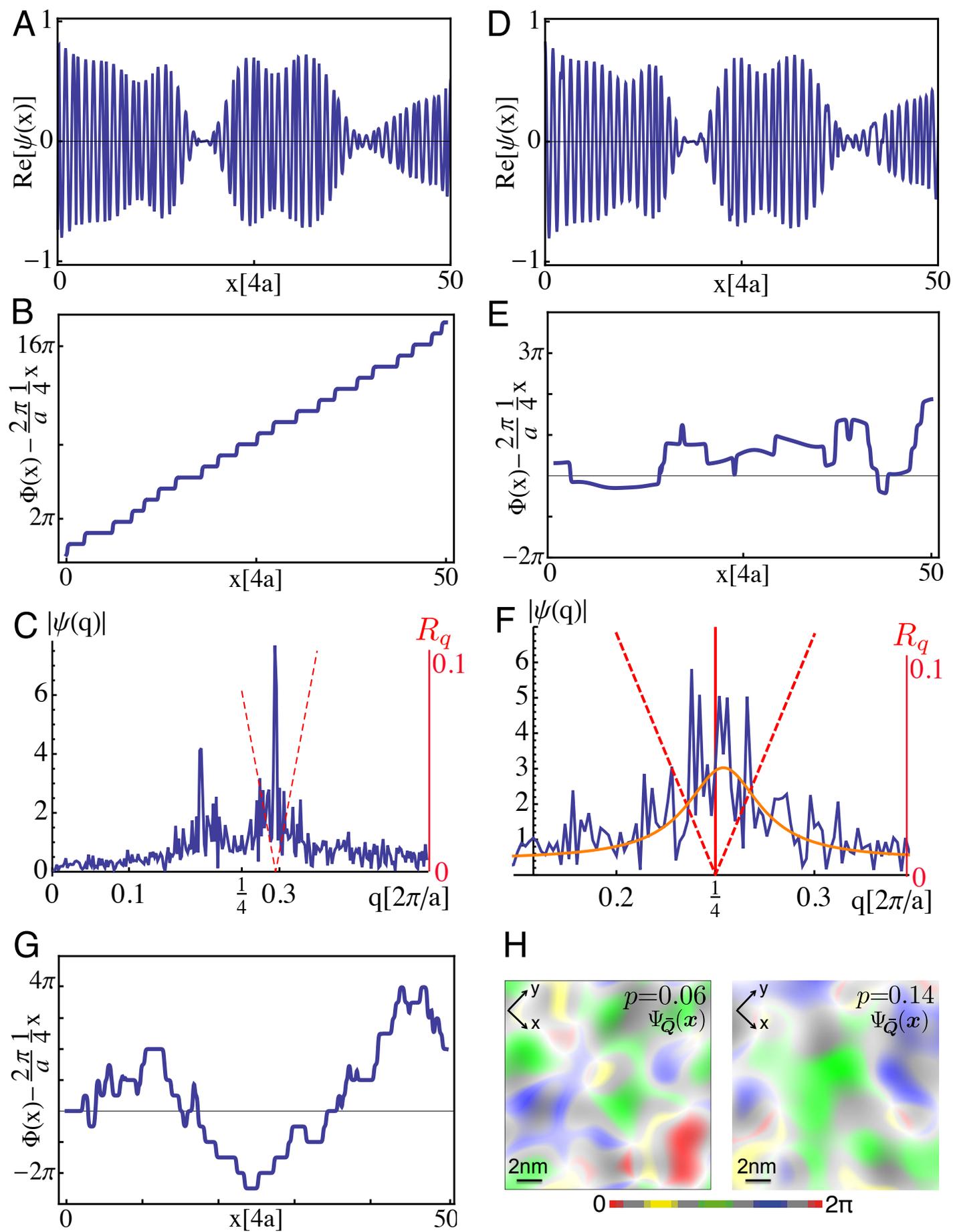
Figure S1

Figure S2

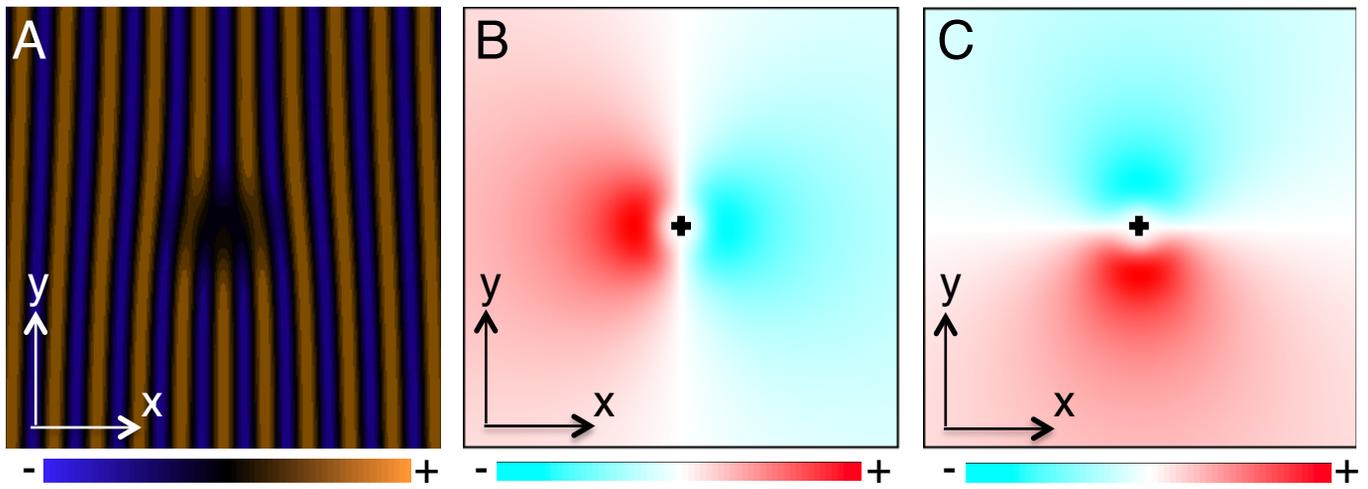

Figure S3

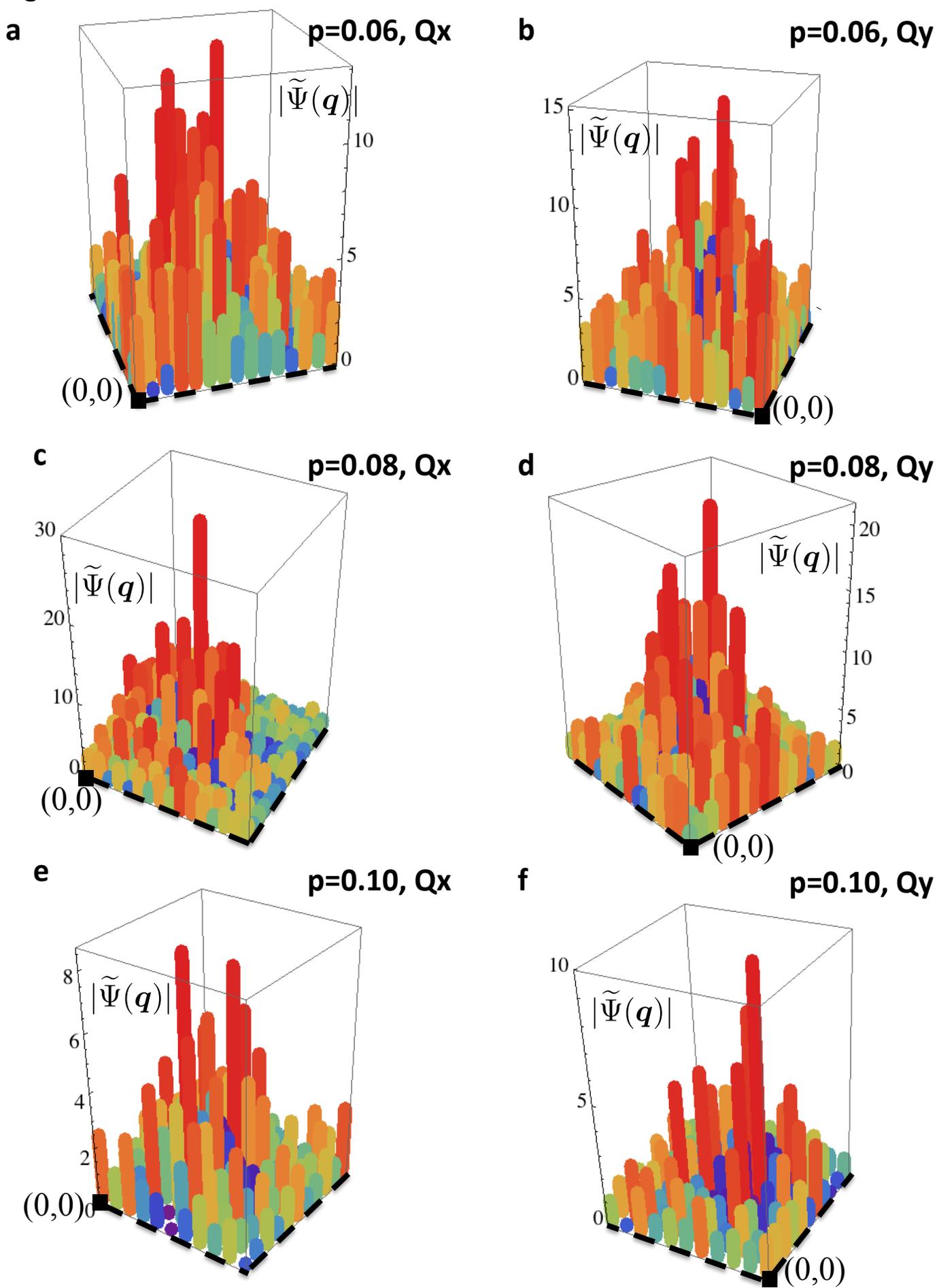

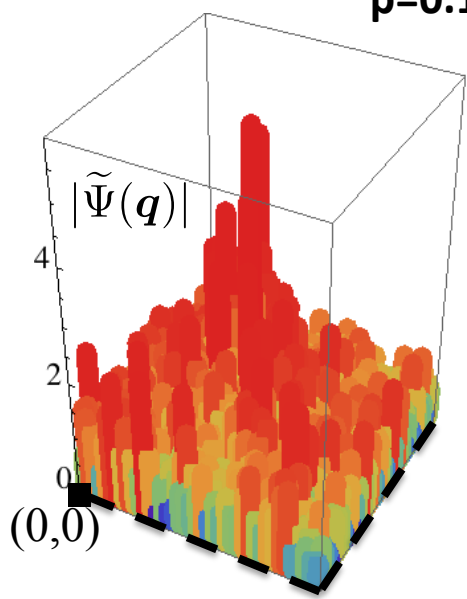

**g** **p=0.14, Qx**

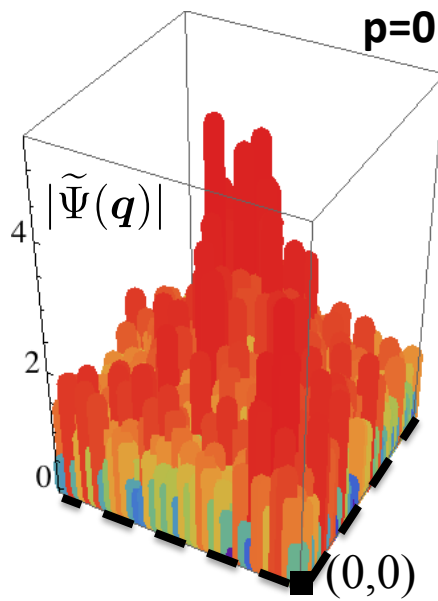

**h** **p=0.14, Qy**

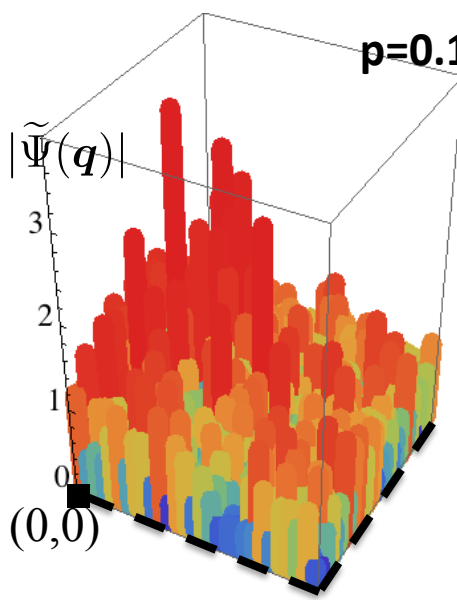

**i** **p=0.17, Qx**

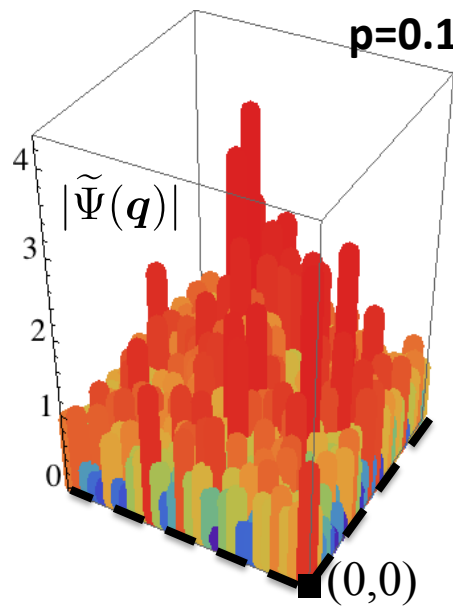

**j** **p=0.17, Qy**

Figure S4

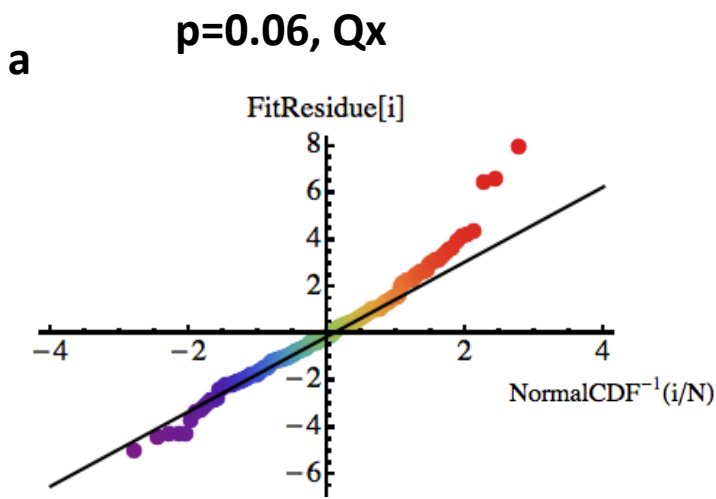
**a** p=0.06, Qx

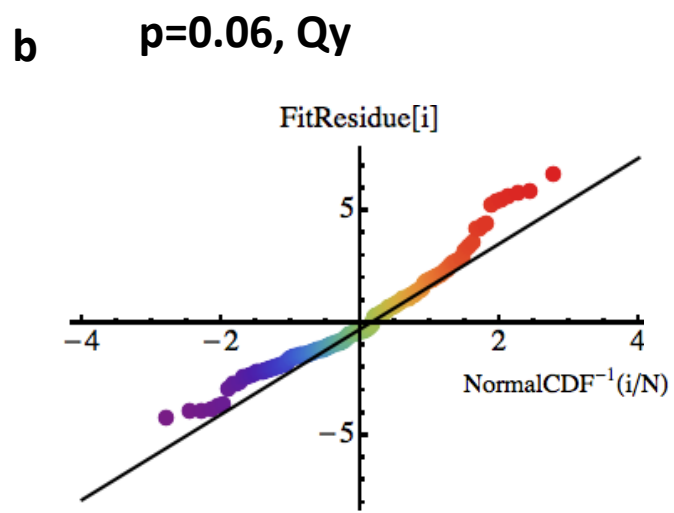
**b** p=0.06, Qy

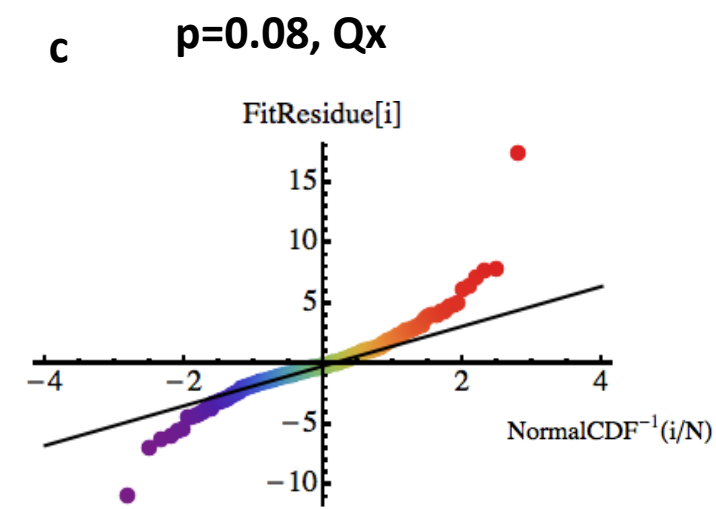
**c** p=0.08, Qx

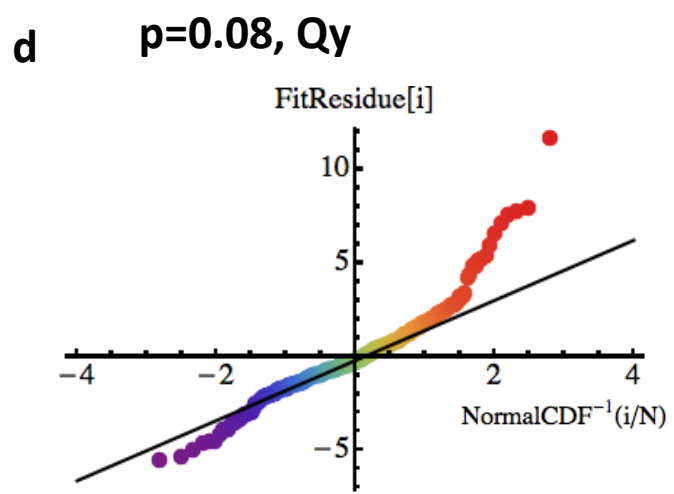
**d** p=0.08, Qy

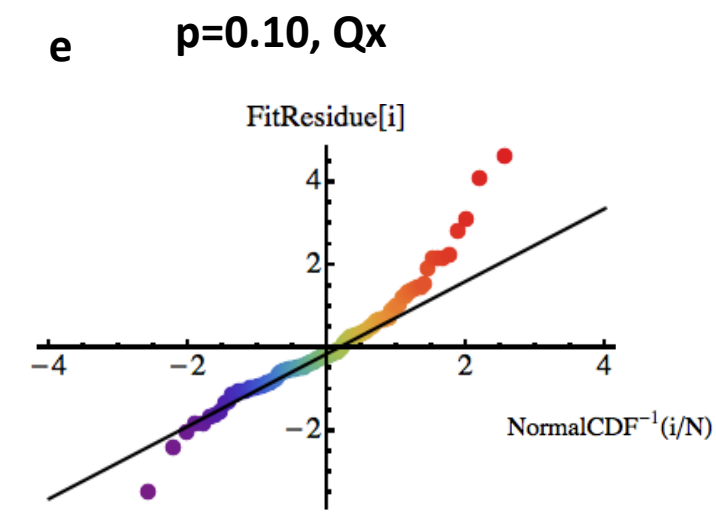
**e** p=0.10, Qx

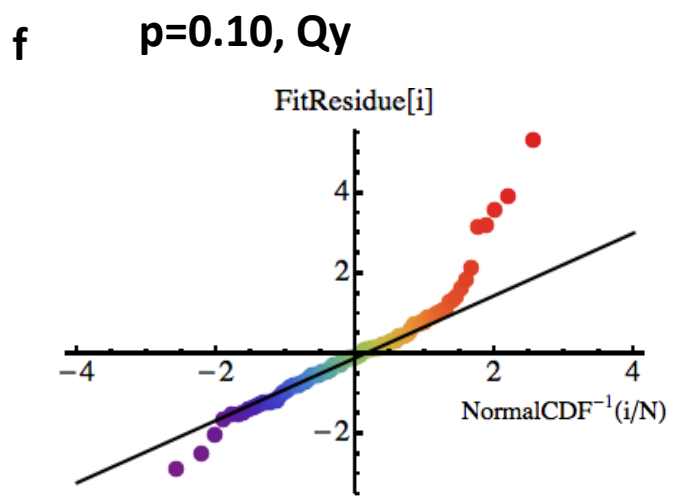
**f** p=0.10, Qy

**g** **p=0.14, Qx**

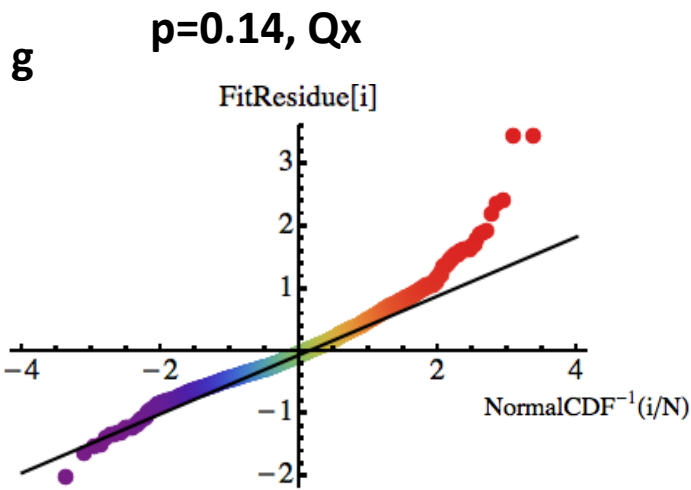

**h** **p=0.14, Qy**

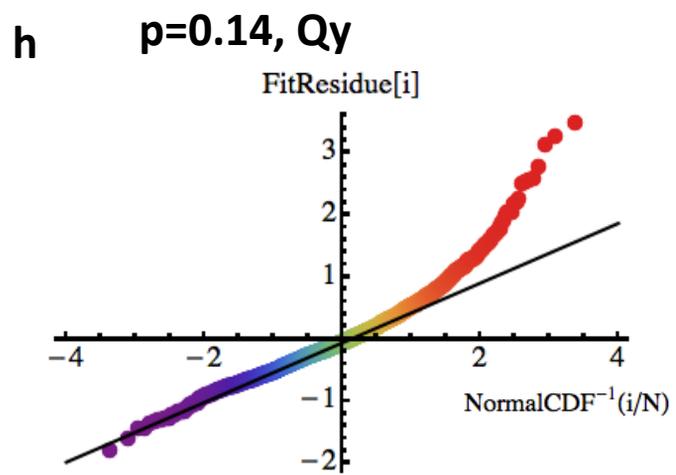

**i** **p=0.17, Qx**

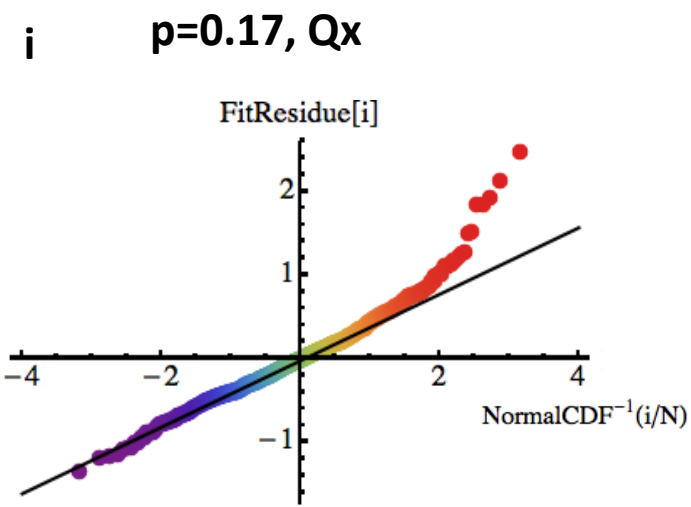

**j** **p=0.17, Qy**

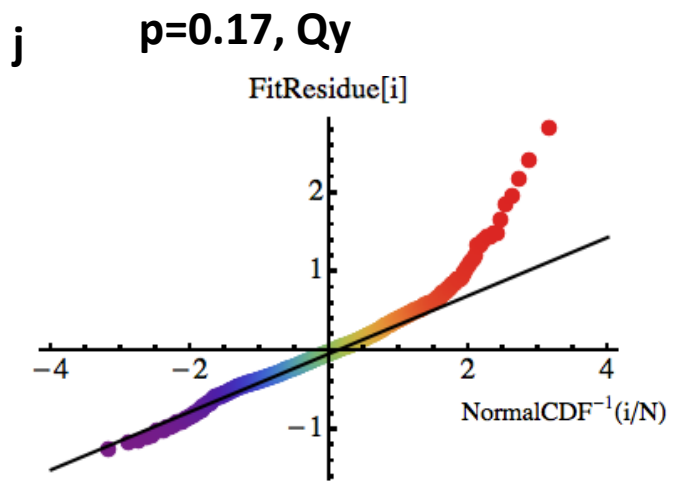

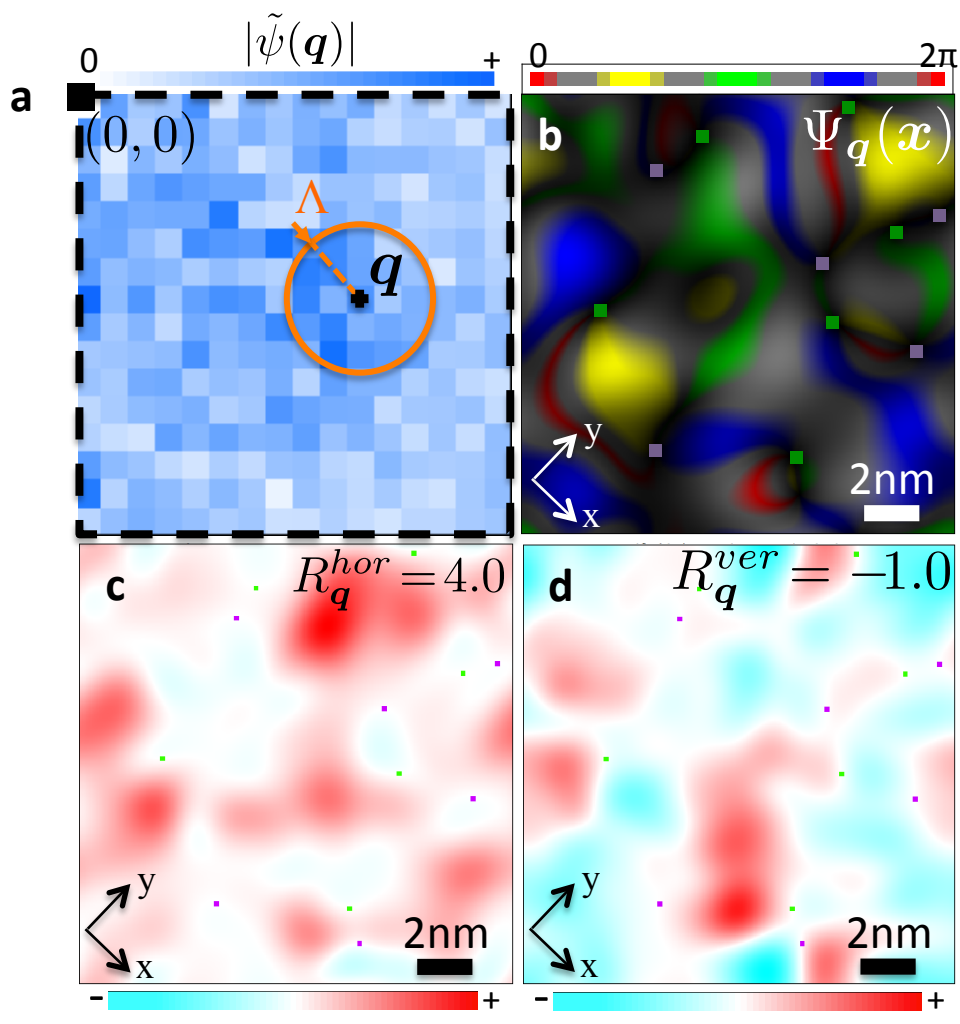

Figure S6

A 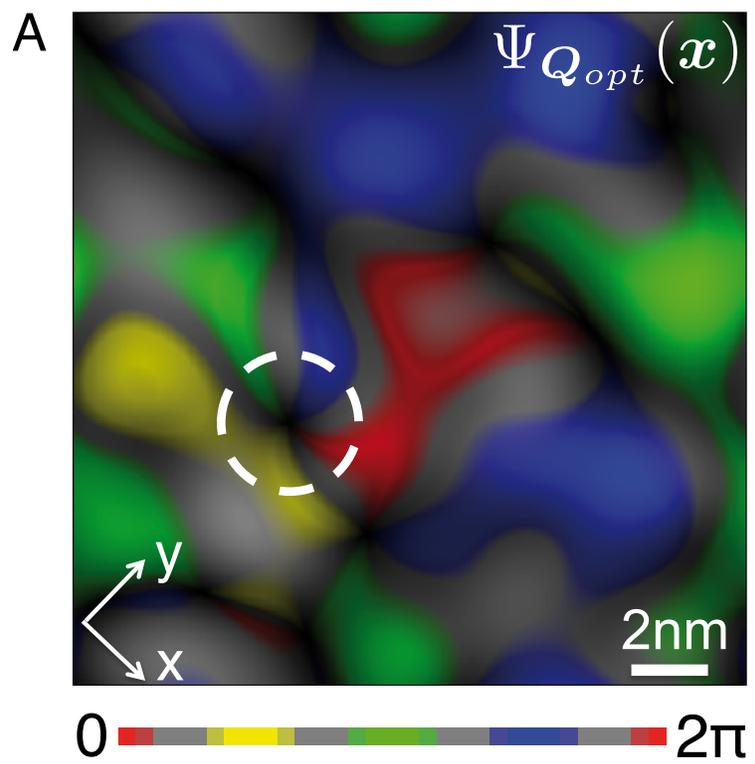

B 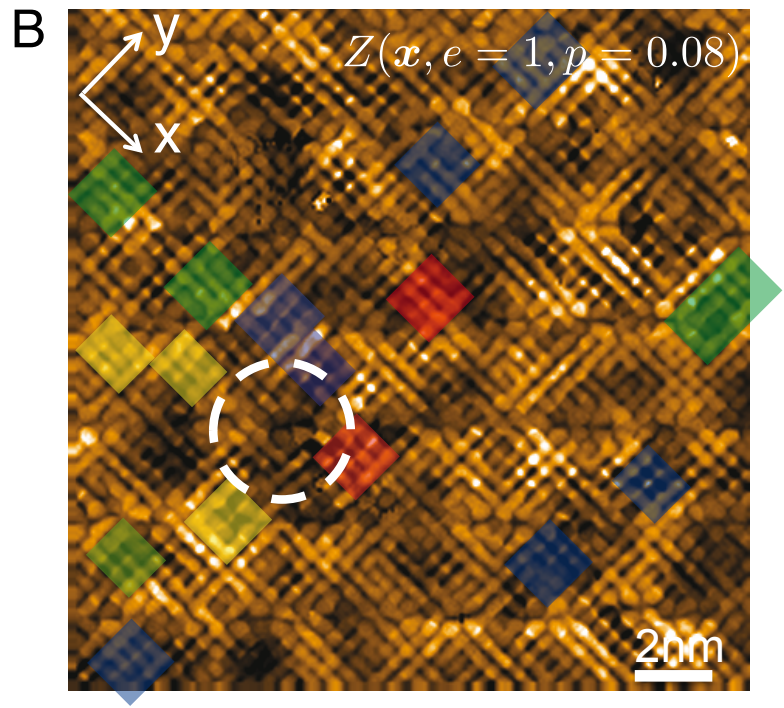

Figure S7

A  $p=0.06, \mathbf{Q}_X$

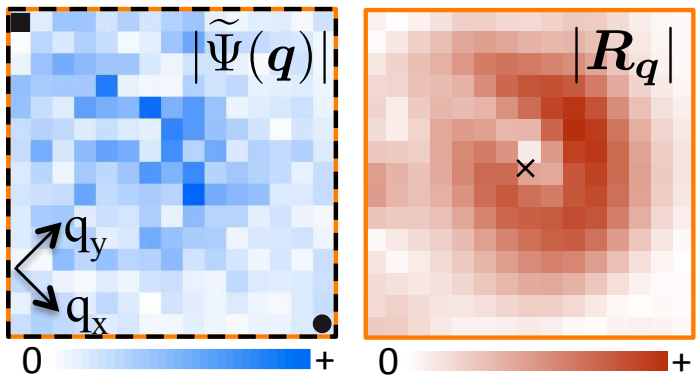

B  $p=0.08$

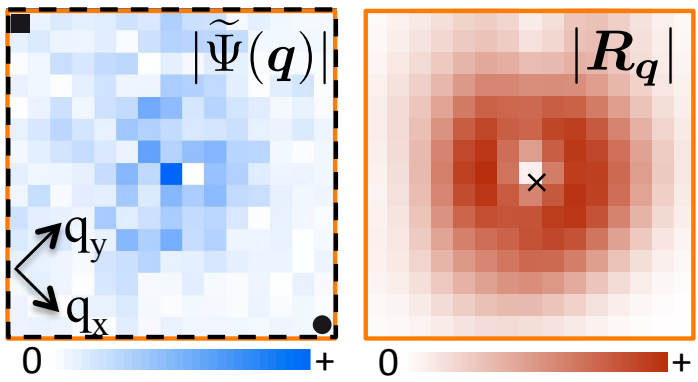

C  $p=0.10$

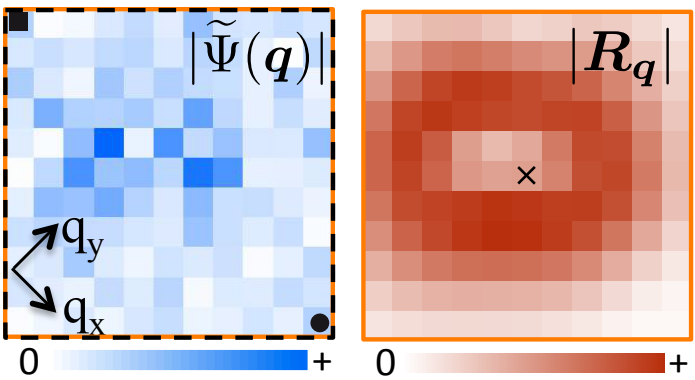

D  $p=0.14$

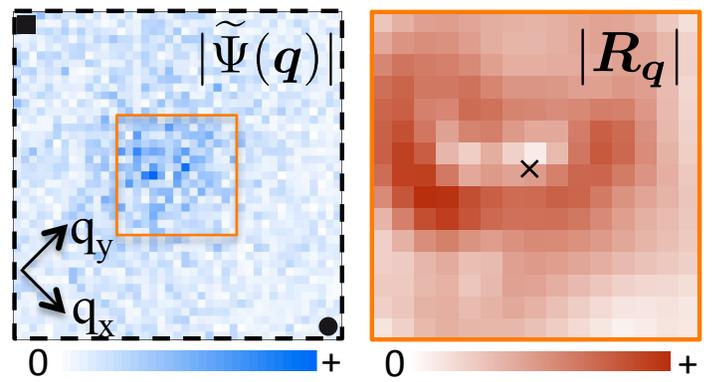

E  $p=0.17$

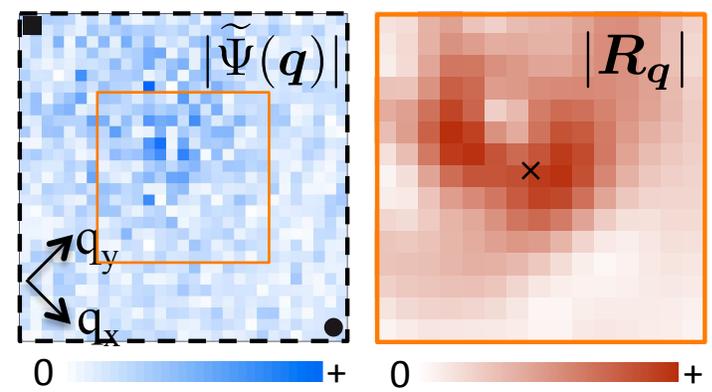

Figure S8

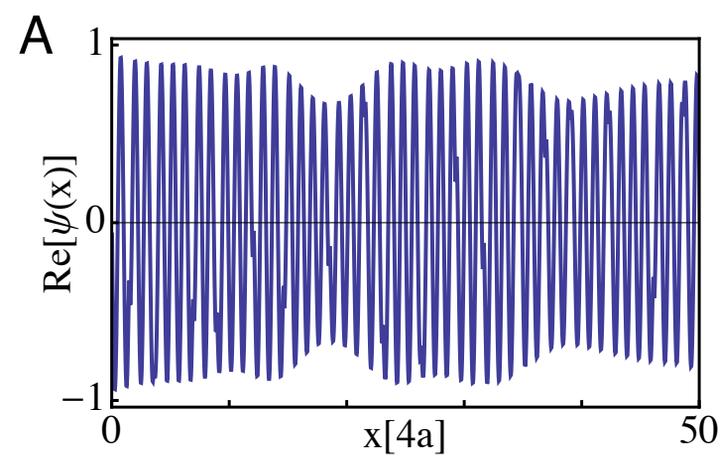

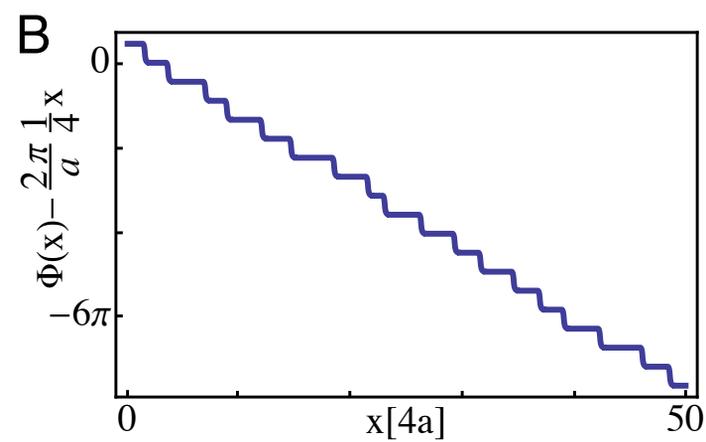

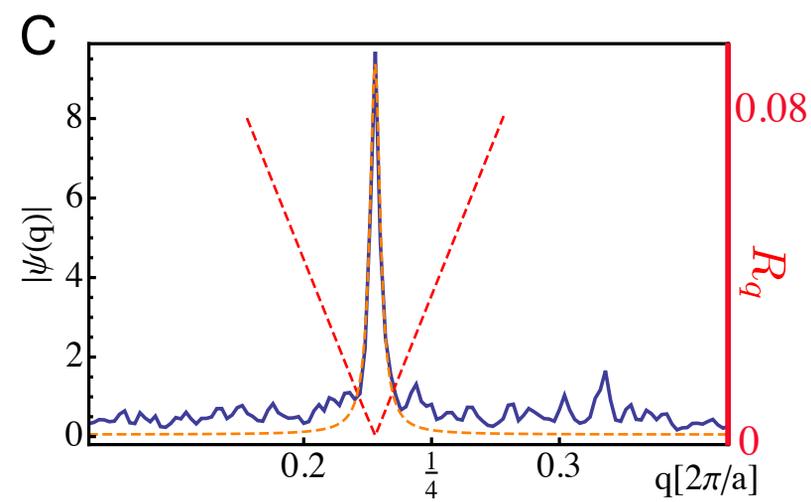